\documentclass[a4paper]{IEEEtran}
\usepackage[T1]{fontenc}
\usepackage[utf8]{inputenc}
\usepackage{amsmath}
\usepackage{paralist}
\usepackage{pgfplots}
\usepackage{subfig}
\usepackage{tikz}
\usepackage{algorithm}
\usepackage{algorithmic}

\usetikzlibrary{calc}
\usetikzlibrary{intersections}
\usetikzlibrary{positioning}
\usetikzlibrary{decorations.markings}
\usetikzlibrary{decorations.shapes}

\tikzstyle{circ} = [shape = circle, draw, inner sep = 1 pt]

\tikzstyle{my dots} = [decorate,
  decoration = {shape backgrounds, shape = circle,
    shape evenly spread = {#1}, shape size = 3 pt}]

\tikzstyle{muwave} = [<->, > = stealth, gray, decorate,
  decoration = {snake, amplitude = 0.25 mm, segment length = 1 mm,
    post length = 1 mm, pre length = 1 mm}]

\newcommand{\BS}[1]{%
  \begin{tikzpicture}[#1]
    \fill (0, 0) circle (5 mm);
    \draw ([shift = (-45:15 mm)] 0, 0) arc (-45:45:15 mm);
    \draw ([shift = (135:15 mm)] 0, 0) arc (135:225:15 mm);
    \draw ([shift = (-45:10 mm)] 0, 0) arc (-45:45:10 mm);
    \draw ([shift = (135:10 mm)] 0, 0) arc (135:225:10 mm);
    \coordinate (ll) at ($ (0, 0) + (-15 mm, -40 mm) $);
    \coordinate (rl) at ($ (0, 0) + (15 mm, -40 mm) $);
    \draw (0, 0) -- (ll) -- ($ (0, 0)!0.5!(rl) $) -- ($ (0, 0)!0.25!(ll) $);
    \draw (0, 0) -- (rl) -- ($ (0, 0)!0.5!(ll) $) -- ($ (0, 0)!0.25!(rl) $);
  \end{tikzpicture}%
}

\newcommand{\DC}[1]{%
  \begin{tikzpicture}[#1]

    \draw (0, 0) ellipse [x radius = 25 mm, y radius = 10 mm];
    \draw (-25 mm, 0) -- (-25 mm, -40 mm);
    \draw (25 mm, 0) -- (25 mm, -40 mm);
    \draw (-25 mm, -40 mm) arc (180:360:25 mm and 10 mm);

  \end{tikzpicture}%
}

\newcommand{\EC}{%
  \begin{tikzpicture}
  \draw [green, dash pattern = on 6 pt off 5 pt, line width = 2.5
    pt, opacity = 0.5, rounded corners = 4 pt, line cap = round]
  (0, 0) -- (0.6, 0);
  \end{tikzpicture}%
}

\newcommand{\CP}{%
  \begin{tikzpicture}
   \fill [red, opacity = 0.5, decorate, decoration = {shape
       backgrounds, shape = circle, shape evenly spread = 5}]
  (0, 0) -- (0.6, 0);
  \end{tikzpicture}%
}

\newcommand{\ST}{%
  \begin{tikzpicture}
     \draw [line width = 2pt]
  (0, 0) -- (0.6, 0);
  \end{tikzpicture}%
}

\newcommand{\usecasenetwork}{%
  \tikzstyle{eonn} = [circle, draw, minimum size = 0.4 cm,
    inner sep = 0]

  \node [circ] (bs1) [label = BS1] at (0, 0)
        {\BS{scale = 0.09}};

  \node [circ] (bs2) [right = of bs1, label = BS2]
        {\BS{scale = 0.09}};

  \tikzset{node distance = 0.7 cm}

    \node [eonn] (eona) [below = of bs1] {A};
    \node [eonn] (eonb) [below = of bs2] {B};
    \node [eonn] (eonc) [below = of eona] {C};
    \node [eonn] (eond) [below = of eonb] {D};

    \node [circ] (dc1) [below = of eonc, label = below:DC1]
          {\DC{scale = 0.08}};
    \node [circ] (dc2) [below = of eond, label = below:DC2]
          {\DC{scale = 0.08}};

    \draw [dashed] (bs1) -- (eona);
    \draw [dashed] (bs2) -- (eonb);

    \draw (eona) -- (eonb);
    \draw (eona) -- (eonc);
    \draw (eonb) -- (eond);
    \draw (eonc) -- (eond);

    \draw (eonc) -- (dc1);
    \draw (eond) -- (dc2);

    \def\dfncb{0.3 cm}
}

\title{Itinerant routing in elastic optical networks}

\author{Ireneusz Szcześniak, Artur Gola, Andrzej Jajszczyk, Andrzej
R.~Pach, and Bożena Woźna-Szcześniak%
  \thanks{This work was supported by the postdoctoral fellowship
    number DEC-2013/08/S/ST7/00576 from the Polish National Science
    Centre.  The simulation results were obtained using PL-Grid, the
    Polish supercomputing infrastructure.  This is the extended
    version of a conference paper \cite{10.1109/ICCC}.}%
  \thanks{I.~Szcześniak is with the Institute of Computer and
    Information Sciences of the Częstochowa University of Technology,
    Częstochowa, Poland.}%
  \thanks{A.~Gola and B.~Woźna-Szcześniak are with the Institute of
    Mathematics and Computer Science of the Jan Długosz University,
    Częstochowa, Poland.}%
  \thanks{A.~Jajszczyk and A.~R.~Pach are with the Department of
    Telecommunications of the AGH University of Science and
    Technology, Krakow, Poland.}}

\begin{document}

\maketitle

\begin{abstract}

  We state a new problem of \emph{itinerant routing} in elastic
  optical networks, which we define as the establishment and
  reconfiguration of an \emph{itinerant connection}.  In an itinerant
  connection, one of the connection end nodes is allowed to change.
  Itinerant routing could also be considered a form of connection
  restoration, where a connection has to be restored to a different
  end node.  With the introduction of new mobile-network
  architectures, the progressing integration of wireless and optical
  networks, the continuing growth of wireless high-bitrate services,
  and the eventual deployment of elastic optical networks, there is a
  need to support this itinerant routing in the elastic optical
  networks.  We present and discuss two major use cases of the
  itinerant routing, and propose a \emph{novel reconfiguration
  algorithm}.  Our algorithm does not deteriorate the network
  performance, and requires half as many new links to configure as the
  shortest-path reconfiguration algorithm.  The performance evaluation
  was carried out with 46800 simulation runs using randomly-generated
  realistic transport networks.

\end{abstract}


%%%%%%%%%%%%%%%%%%%%%%%%%%%%%%%%%%%%%%%%%%%%%%%%%%%%%%%%%%%%%%%%%%%%%%%%%%%

\section{Introduction}

% Wireless networks will grow.  Network densification is key.

The wireless access network traffic has increased manyfold in recent
years along with the number of users, and network operators are
expecting its further dynamic increase which has to be serviced by the
optical transport network \cite{10.1109/JLT.2015.2394297}.  Since this
trend is expected to continue, the network densification across all
network layers is regarded as a key and inevitable enabler of the
fifth generation mobile (5G) networks
\cite{10.1109/MCOM.2014.6736747}, putting higher requirements on the
optical transport network.

% The importance of the optical networks and OTN in the MAN.

Network densification is a major reason why optical transport networks
should continue to grow in the metropolitan area networks (MANs).
Currently the Optical Transport Network (OTN) is deployed in MANs over
wavelength-division multiplexing (WDM) networks \cite{lavacca}.  Even
though OTN was designed for long-haul communication, it is also
deployed in MANs because it can economically switch a large number of
subwavelength short-distance circuits.

% The future optical networks should be elastic.

The elastic optical networks (EONs) are very likely to be deployed
because they utilize the limited erbium window more efficiently than
the incumbent WDM networks \cite{10.1109/MCOM.2012.6146481}.  The
flex-grid has now been standardized with 6.25 GHz channel spacing
\cite{ITU-T/G.694.1}, but as the technology is deployed and further
improved, we can expect smaller channel spacing in the future,
supporting lower client bitrates, which would increase network
utilization, lower blocking probability and lower spectrum
fragmentation.  For even lower bitrates, OTN over the EONs is an
option \cite{6476195}.

% C-RAN: the future of 5G networks.  BSs connect to DCs.

The centralized radio access network (C-RAN, also called the cloud
RAN) is a major candidate for a 5G network architecture
\cite{10.1109/MCOM.2014.6957145}.  In the C-RAN, the base station (BS)
is kept simple to allow for a large-scale deployment of inexpensive
small cells (picocells or femtocells), while the data processing is
relegated to various data centers (DCs), which allows for the load
balancing of radio signal processing, the network function
virtualization (NFV), and easier hardware upgrades.  However, the
C-RAN requires agile high-bitrate connections between BSs and DCs for
transporting digitized radio signals according to bandwidth-demanding
standards like, most notably, the Common Public Radio Interface
(CPRI).  The ambitious operation of the C-RAN, and quite likely of
other 5G architectures, is termed the \emph{wireless-optical network
orchestration}, and is being enabled by the first commercial silicon
photonic integrated circuits \cite{10.1109/JLT.2015.2394297}.  The
wireless-optical network orchestration requires the support of
itinerant routing in the EONs.

% Mobile high bitrate services.

There are a number of novel 5G use cases, where the capacity provided
by the transport network should migrate between BSs to efficiently
utilize the transport capacity.  For example, to provide service in
buses and trains, the moving femtocell (MFC)
\cite{10.1109/MCOM.2014.6736752} migrates between BSs, and along with
it the capacity should migrate too.  Likewise, moving crowds and the
emergency vehicles also require the capacity to migrate between BSs.
These mobile high-bitrate wireless services require the support of
itinerant routing in the EONs.

% Our contribution and what we study specifically.

Our contribution is the statement of the new research problem of
itinerant routing in the EONs, and its solution: a novel
reconfiguration algorithm.  We study the algorithm performance in
comparison with the shortest-path reconfiguration.  This paper is an
extension of a conference paper \cite{10.1109/ICCC}, where we
introduced the itinerant routing, though we initially called it the
mobile routing.  However, we renamed it to itinerant routing, not to
confuse it with routing in mobile networks.

% On our simulations.

We carry out the simulative performance evaluation of the
reconfiguration algorithms with three algorithms for finding candidate
paths, and three policies of allocating spectrum.  The two metrics of
chief interest compared are the number of new links to configure, and
the probability of reconfiguring a connection, along with a number of
other metrics which augment the comparison.

% Paper organization.

The paper is organized as follows.  In the following section we review
key related works.  In Section \ref{statement} we state the research
problem, in Section \ref{algorithm} we describe the proposed
algorithm, and in Section \ref{results} we report the obtained
performance evaluation results.  Section \ref{conclusion} concludes
the paper.

%%%%%%%%%%%%%%%%%%%%%%%%%%%%%%%%%%%%%%%%%%%%%%%%%%%%%%%%%%%%%%%%%%%%%%%%%%%

\section{Related works}
\label{related}

% How we improved the study.

This paper differs from \cite{10.1109/ICCC} in a number of ways.
First, we extend the study of itinerant routing, so that candidate
paths are computed not only with the adapted and constrained Dijkstra
algorithm for the EONs \cite{10.1109/ONDM.2016.7494087}, but also with
two other algorithms frequently used: the Yen K-shortest paths
(Yen-KSP) algorithm, and the link-disjoint all shortest paths (LD-ASP)
algorithm.  Second, we replace a complicated traffic model with a
simple and tractable one.  Next, we extended the study to include the
random spectrum allocation policy.  Finally, we remove the incremental
reconfiguration algorithm from the study, since we have already shown
it is markedly inferior, and since its performance results would only
obfuscate the study.

% Routing in EONs.

Reconfiguration in itinerant routing is a form of routing and spectrum
assignment (RSA) in EONs.  Several variations of the RSA problem in
EONs have been well studied, both for many demands at once and for a
single demand using the integer linear programming and heuristic
algorithms \cite{10.1109/JLT.2014.2315041, 10.1016/j.osn.2014.02.003}.
The RSA problem has been tackled in various contexts, e.g., grooming
or virtual network embedding, and to the best of our knowledge, the
end nodes of a demand in these problems do not change, while in our
case they do.

% On GoS and our contribution to it.

In \cite{10.1109/MCOM.2007.313400} the authors argue that the quality
of service (QoS) of optical connections does not take into account the
parameters of the connection setup, rerouting or tear down, such as
the preemption probability, and therefore introduce the grade of
service (GoS) for optical connections to address these parameters.
Our work contributes new results on the GoS of itinerant routing in
elastic optical networks.

% Mobile backhaul.

The research on mobile backhaul usually concentrates on PONs which
integrate the wireless access network with the optical transport
network \cite{10.1109/JLT.2010.2050861}.  We, however, concentrate on
the support of itinerant routing in the EONs required by the mobile
backhaul, which will further gain in importance if the mobile backhaul
becomes elastic.

% Moving extended cell.

In \cite{10.1109/JLT.2009.2022505} the authors propose the concept of
the moving extended cell (MEC), which guarantees the high-bitrate
connectivity for a highly mobile client by transceiving wireless
signal for this client also in the picocells neighboring the picocell
in which the client currently resides.  Our algorithm can be used to
reconfigure quickly a high-bitrate connection when the client roams
between picocells serviced by different hotels.

% Radio over fiber.

Our research results are well suited for radio-over-fiber (RoF) mobile
backhaul which amalgamates the optical and wireless networks by
transceiving subcarriers \cite{10.1109/SURV.2013.013013.00135}.  In
\cite{10.1109/JLT.2013.2286564} the authors propose a cost-effective
RoF architecture which converts the 60 GHz radio subcarriers directly
into the subcarriers of the optical erbium window, thus enabling the
elastic optical networks to make inroads into mobile backhaul.

% On anycast routing.

We note that itinerant routing is not the anycast routing, though it
bears some resemblance.  In the anycast routing, commonly used in the
Internet, the destination node of a connection can be any from a given
set, but once the connection is established, the destination node does
not change.  The hallmark of the itinerant routing is that the end
nodes can change during the duration of the connection.

%%%%%%%%%%%%%%%%%%%%%%%%%%%%%%%%%%%%%%%%%%%%%%%%%%%%%%%%%%%%%%%%%%%%%%%%%%%

\section{Problem statement}
\label{statement}

% A short introduction to EONs.

We model the EONs by an undirected graph $G(V, E)$, where $V$ is the
set of nodes, and $E$ is the set of links.  The links have the
attributes of length and available slices.  Optical spectrum is
divided into slices of equal bandwidth, with $\Omega$ being the set of
all slices.  A given number of contiguous slices create a slot, which
can be used to transport an optical signal.  A bidirectional unicast
symmetric connection is established by allocating the same slices on
undirected links.

% On itinerant connections.

We introduce a new concept of the itinerant connection, where one of
the connection end nodes is allowed to change during the duration of
the connection.  A connection is established between the source node
$s$ and the destination node $t$ with the set of contiguous slices $S$
along the path $P$.  The problem is to find a reconfiguration path
$P'$, and a set of reconfiguration slices $S'$, when the connection
has to be reconfigured for the new destination node $t'$.  The problem
of reconfiguring the connection for a new source node $s'$ is the
same.

% Two use cases, and the relation to the figures.

\begin{figure}
  \centering
  \subfloat[before reconfiguration]{\label{uc:before}\input{usecase1}}
  \hfill
  \subfloat[after reconfiguration]{\label{uc:after}\input{usecase2}}
  \caption[Two use cases]{Two use cases of itinerant routing, where
    \BS{scale = 0.06} is a base station, \DC{scale = 0.05} is a data
    center, \includegraphics[scale=0.035]{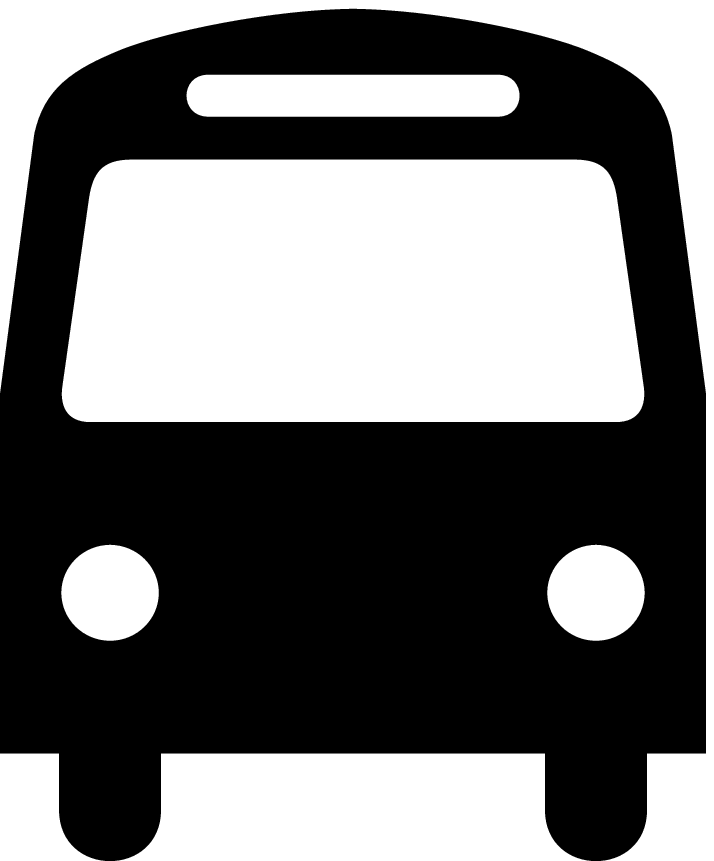} is a means of public
    transport, and A, B, C, and D are the EON nodes.}
  \label{usecases}
\end{figure}

There are two major use cases of an itinerant connection in 5G networks:
\begin{inparaenum}
\item a connection is switched to a different DC, and
\item a connection is switched to a different BS.
\end{inparaenum}  These use cases are illustrated in
Fig.~\ref{usecases}, where the connections before the reconfiguration
are shown in Fig.~\ref{uc:before}, and after in Fig.~\ref{uc:after}.

% The first use case.

In the first use case, a connection from a BS is switched to a new DC,
because the current DC is overloaded or went down, or the new DC
offers a better deal.  As shown in Fig.~\ref{usecases}, the connection
from BS1 is switched from DC1 to DC2.

% The second use case.

In the second use case, a connection from a DC is switched to a new
BS, because an MFC or a crowd roams to this BS.  As shown in
Fig.~\ref{usecases}, the MFC of a bus makes a transition from BS2 to
BS1, and the transport capacity migrates along, i.e., the connection
from DC2 is switched from BS2 to BS1.

% General objective of reconfiguration.

In general, the problem is how to reconfigure the connection fast and
frugally.  We want to minimize both the reconfiguration time and the
resources used, but there is a trade-off between them in that the
faster we want to reconfigure the connection, the more resources we
might need to spend.

%%%%%%%%%%%%%%%%%%%%%%%%%%%%%%%%%%%%%%%%%%%%%%%%%%%%%%%%%%%%%%%%%%%%%%%%%%%

\section{Proposed algorithm}
\label{algorithm}

% What we concentrate on: the number of new links.

Specifically, we concentrate on the number of link configurations,
since they require substantial time, and can cause service
disruptions.  Limiting the number of the required link configurations
would also limit the control traffic in the network, and limit the
network destabilization.

% Complete reconfiguration.

A natural solution to the problem is to tear down the established
connection, and search for a new path from the source node to the new
destination node $t'$ with any available slices capable of supporting
a demand.  We call it the \emph{complete reconfiguration}.

% How the proposed algorithm achieves its goal: bridge, reuse.

However, we propose a reconfiguration algorithm, which minimizes the
number of the required new links by reusing the links of the already
established connection.  The algorithm attempts to optimally
\emph{bridge} one of the nodes of the already established connection
to the new destination node.  The algorithm could be implemented in the
path computation element (PCE) of the Generalized Multiprotocol Label
Switching (GMPLS) control plane.

% How to find the bridging and candidate reconfiguration paths.

For every node $n$ of the path $P$ of the established connection, the
algorithm searches for a shortest \emph{bridging path} $B_n$ from a
node $n$ to the new destination node $t'$, provided the bridging path
meets the \emph{spectrum contiguity and continuity constraints}, i.e.,
it has to use on all its links the same set of contiguous slices $S$
already used by the established connection.  Having a bridging path
found, a \emph{candidate reconfiguration path} $P_n$ is produced by
concatenating $C_n$ and $B_n$, where $C_n = \text{CUT}(P, n)$ is the
front of the path $P$ cut at a node $n$, as returned by the function
$\text{CUT}(P, n)$.

% Candidate paths, priority queue, and complete reconfiguration.

Candidate reconfiguration paths are pushed into the priority queue
$Q$, where they are sorted in ascending order first by the number of
the new links they require (i.e., by the number of links in $B_n$),
and next by the number of all their links, new or reused (i.e., by the
number of links in $P_n$).  The number of candidate reconfiguration
paths pushed into $Q$ equals to at most the number of nodes in the
path $P$.  Once all the candidate paths have been pushed into $Q$, the
solution sought is at the front of the queue.  If no candidate
reconfiguration path is found with bridging, the algorithm resorts to
the complete reconfiguration.

% The bridging paths belong to the reverse shortest-path tree.

The shortest bridging paths belong to the reverse shortest-path tree
(SPT) rooted at the new destination node $t'$.  The paths in a reverse
SPT end at the root, and lead from other nodes, as opposed to an SPT
whose paths start at the root, and lead to other nodes.  We get a
reverse SPT by reversing an SPT obtained for $G^R$, the reverse graph
of $G$.  For an undirected graph $G$, an SPT and its reverse SPT are
the same, but we keep the distinction anyway to make the algorithm
applicable to directed graphs too.

% How we grow the reverse SPT.

We grow an SPT using our shortest-path algorithm for the EONs
\cite{10.1109/ONDM.2016.7494087}: an adapted and constrained Dijkstra,
which takes into account the spectrum continuity and contiguity
constraints.  The algorithm finds all available slices in the SPT.
The objective of the algorithm is to find a shortest path to a given
node (or a given set of nodes), and to this end the algorithm builds
an SPT.  To grow a minimal SPT, the algorithm stops when all the nodes
of the established connection are reached.

% The algorithm description in general.

Algorithm \ref{algorithm} displays the proposed reconfiguration
algorithm.  The algorithm takes as input the graph $G$, the path $P$,
the set of slices $S$, the source node $s$, and the new destination
node $t'$, and produces the reconfiguration path $P'$ and the
reconfiguration set of slices $S'$.  If no solution can be found, $P'$
and $S'$ are returned null.

% The workings of the algorithm.

The algorithm produces the bridging paths $B_n$ from the reverse SPT
$T$, obtained by reversing the SPT returned by the SPT function.  The
function $\text{PATH }(T, S, n)$ returns a path from a node $n$ in the
reverse SPT $T$, provided the path can support exactly the set of
slices $S$.  The function $\text{PATHSSC }(T, s, |S|)$ returns a pair
of a path from the node $s$ in the reverse SPT $T$, and a set of $|S|$
contiguous slices, where $|S|$ is the cardinality of $S$.  The
returned set of contiguous slices is selected from the set of slices
available along that path according to an employed spectrum selection
policy.

\begin{algorithm}[t]
  \caption{\\
    {\bf{}Input}: $G$, $P$, $S$, $s$, $t'$\\
    {\bf{}Output}: $P', S'$}
  \label{algorithm}
  \begin{algorithmic}
    \STATE $Q \leftarrow \emptyset$ \COMMENT{priority queue of candidate reconfiguration paths}
    \STATE $T = \text{SPT }(G^R, t', \text{nodes of } P)^R$
    \FORALL{$n \in \text{nodes of }P$}
    \STATE $B_n = \text{PATH }(T, S, n)$
    \IF{$B_n$ exists}
    \STATE $C_n = \text{CUT }(P, n)$
    \STATE $P_n = C_n + B_n$
    \STATE push $P_n$ to $Q$
    \ENDIF
    \ENDFOR
    \IF{$Q$ is not empty}
    \STATE $P' = \text{front of } Q$
    \STATE $S' = S$
    \ELSE
    \STATE $P', S' = \text{PATHSSC }(T, \Omega, s, |S|)$
    \ENDIF
    \RETURN $P', S'$
  \end{algorithmic}
\end{algorithm}

% Description of the example figure.

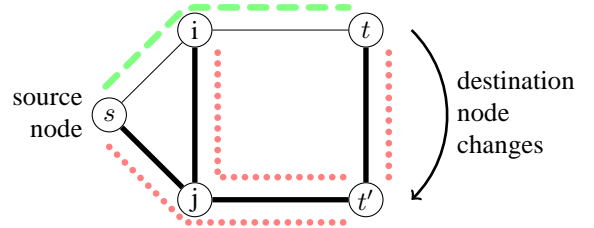
\begin{figure}
  \centering
  \begin{tikzpicture}[scale = 1.5]

  \tikzstyle{eonn} = [circle, draw, minimum size = 0.45 cm,
    inner sep = 0]

  \node [eonn] (n1) at (0.25, 0) {$s$};
  \node [eonn] (n2) at (1, 0.75) {i};
  \node [eonn] (n3) at (2.5, 0.75) {$t$};
  \node [eonn] (n4) at (1, -0.75) {j};
  \node [eonn] (n5) at (2.5, -0.75) {$t'$};

  \draw [line width = 0.25 pt] (n1) -- (n2);
  \draw [line width = 2 pt] (n1) -- (n4);
  \draw [line width = 0.25 pt] (n2) -- (n3);
  \draw [line width = 2 pt] (n2) -- (n4);
  \draw [line width = 2 pt] (n3) -- (n5);
  \draw [line width = 2 pt] (n4) -- (n5);

  \def\dfnca{0.2 cm}

  % The red path --------------------------------------
  \coordinate (p1a) at ($ (n1.45) + (135:\dfnca) $);
  \coordinate (p1b) at ($ (n2) + (135:\dfnca) $);
  \coordinate (p1c) at ($ (n2) + (90:\dfnca) $);
  \coordinate (p1d) at ($ (n3.180) + (90:\dfnca) $);

  \coordinate (c1a) at (intersection of p1a--p1b and p1c--p1d);

  \draw [green, dash pattern = on 6 pt off 5 pt, line width = 2.5
    pt, opacity = 0.5, rounded corners = 4 pt, line cap = round]
  (p1a) -- (c1a) -- (p1d);

  \def\dfnca{0.2 cm}
  \def\dfncb{0.2 cm}
  \def\dfncc{0.4 cm}

  % The 1st green path --------------------------------------
  \coordinate (p2a) at ($ (n1) + (-45:\dfncb) + (-135:\dfnca) $);
  \coordinate (p2b) at ($ (n4.135) + (-135:\dfnca) $);
  \coordinate (p2c) at ($ (n4) + (-90:\dfnca) $);
  \coordinate (p2d) at ($ (n5) + (180:\dfncb) + (-90:\dfnca) $);

  \coordinate (c2) at (intersection of p2a--p2b and p2c--p2d);

  \fill [red, opacity = 0.5, decorate, decoration = {shape
      backgrounds, shape = circle, shape evenly spread = 25}]
  (p2a) -- (c2) -- (p2d);

  % The 2nd green path --------------------------------------
  \coordinate (p3a) at ($ (n2) + (-90:\dfncb) + (0:\dfnca) $);
  \coordinate (p3b) at ($ (n4) + (0:\dfnca) $);
  \coordinate (p3c) at ($ (n4) + (90:\dfnca) $);
  \coordinate (p3d) at ($ (n5) + (180:\dfncb) + (90:\dfnca) $);

  \coordinate (c3) at (intersection of p3a--p3b and p3c--p3d);

  \fill [red, opacity = 0.5, decorate, decoration = {shape
      backgrounds, shape = circle, shape evenly spread = 24}]
  (p3a) -- (c3) -- (p3d);

  % The 3th green path --------------------------------------
  \coordinate (p4a) at ($ (n3) + (-90:\dfncb) + (0:\dfnca) $);
  \coordinate (p4b) at ($ (n5) + (90:\dfncb) + (0:\dfnca) $);

  \fill [red, opacity = 0.5, decorate, decoration = {shape
      backgrounds, shape = circle, shape evenly spread = 12}]
  (p4a) -- (p4b);

  % Texts ---------------------------------------------------
  \draw [line width = 1 pt] ($ (n3) + (0:\dfncc) $)
  edge [->, bend left = 45]
  node [right, align = left] {destination\\node\\changes}
  ($ (n5) + (0:\dfncc) $);

  \node [align = right] [left = 0 cm of n1] {source\\node};

\end{tikzpicture}
  \caption[Reconfiguration example]{Reconfiguration example, where
  \raisebox{1.5pt}{\EC} is the established connection,
  \raisebox{1.5pt}{\ST} are the links of the shortest-path tree, and
  \raisebox{1.5pt}{\CP} are the candidate bridging paths rooted in
  $t'$.}
  \label{example}
\end{figure}

Figure \ref{example} shows a reconfiguration example.  A connection
(\raisebox{1.5pt}{\EC}) is established from $s$ to $t$, and the
destination node changes to $t'$.  A reverse SPT is built
(\raisebox{1.5pt}{\ST}), and candidate bridging paths
(\raisebox{1.5pt}{\CP}) are produced.

As reported in \cite{10.1109/ONDM.2016.7494087}, the optimal
shortest-path search can take even a second, which is too long for a
network to react in real time.  Therefore, we propose that the
reconfiguration paths are calculated in advance for the foreseeable
reconfigurations.

%%%%%%%%%%%%%%%%%%%%%%%%%%%%%%%%%%%%%%%%%%%%%%%%%%%%%%%%%%%%%%%%%%%%%%%%%%%

\section{Simulative study}
\label{study}

% What's the deal with the simulations.

We study with simulations the performance of the proposed algorithm in
comparison with the complete reconfiguration.  We are mainly
interested in the number of new links required during the connection
reconfiguration, but we also mind other performance metrics, to ensure
the proposed algorithm does not impede the network performance in
general.  The performance of the two reconfiguration algorithms is
evaluated for various loads, and compared for three routing
algorithms, and three spectrum allocation policies.  We do not use a
single network topology, like NSFNet, but instead a hundred of
randomly-generated realistic network topologies.

% On routing in general.

A routing algorithm is responsible for finding a path for a given
demand.  In our simulations, a routing algorithm is used in three
cases:
\begin{inparaenum}
\item to establish a connection,
\item to find a path for the complete reconfiguration, and
\item to find bridging paths.
\end{inparaenum}

% What routing algorithms we use.

The three routing algorithms compared are as follows: optimal,
Yen-KSP, and LD-ASP.  The optimal routing algorithm finds a
shortest-path (using the algorithm in
\cite{10.1109/ONDM.2016.7494087}), while the Yen-KSP and LD-ASP
algorithms do not guarantee that the path they find is the shortest,
since they are heuristic algorithms.  These routing algorithms search
for a path with the largest set of slices, capable of supporting at
least a given demand.  From that largest set of slices, we select the
contiguous slices according to the spectrum allocation policy.  For a
given generated graph, in the routing algorithms we do not consider
paths which are longer than twice the length of the longest
shortest-path in the graph.

% On Yen-KSP and LD-ASP.

The Yen-KSP routing algorithm finds at most K-shortest paths with the
Yen algorithm, and then selects the shortest one which can support a
given demand.  We search for $K = 10$ paths.  The LD-ASP routing
algorithm finds all link-disjoint paths by repeatedly searching for a
shortest path with the links used by previous shortest paths disabled
until no path can be found.  From among the paths found, the shortest
one capable of supporting a demand is selected.

% The proposed reconfiguration algorithm and the routing algorithms.

In the simulations we also evaluate the performance of the
reconfiguration algorithms when the Yen-KSP and LD-ASP algorithms are
harnessed.  The proposed reconfiguration algorithm efficiently gets
all the bridging paths from a reverse SPT obtained once with the
optimal routing algorithm, but it can also get the bridging paths
separately with the Yen-KSP and LD-ASP algorithms.

% On the spectrum selection policies.

From the set of slots found, we choose exactly the number of requested
contiguous slices according to one of three policies: \emph{first},
\emph{fittest}, or \emph{random}.  In the first available policy, the
slices with the lowest number are chosen.  In the fittest available
policy, the slot with the lowest number of slices are chosen which
still can handle the demand.  In the random policy, the slices are
chosen at random.

% On the client behaviour.

The demand arrival time follows the exponential distribution with the
mean of $\lambda$ hours.  The end nodes of a demand are different and
chosen at random.  A connection is reconfigured only once after it is
established.  The holding time follows the exponential distribution
with the mean of $\beta$ hours, after which the connection is torn
down.  We set the mean holding time $\beta = 10$ hours.

% On the number of slices.

Furthermore, the number of slices of a demand is constant throughout
the duration of the connection and follows the distribution of
$(Poisson(\gamma - 1) + 1)$ with the mean of $\gamma$, i.e., the
Poisson distribution shifted by one to the right, to avoid getting
zero for the number of slices.  We set $\gamma = 10$.

% How a new destination node is selected.

The new destination node $t'$ is chosen at random in such a way that
the number of hops it is away from the destination node $t$ follows
the $(Poisson(0.5) + 1)$ distribution.  We add one to the Poisson
distribution to avoid getting zero, which would result in the
destination node $t' = t$.  This way of choosing $t'$ allows us to
model the connections being frequently switched to a new BS (which are
close to the node $t$), and infrequently switched to a new DC (which
are farther from the node $t$).

% Offered load, and lambda.

We define the offered load $\mu$ as the ratio of the number of slices
demanded to the number of slices in the network.  We vary the offered
load $\mu$ with 26 values of $0.1, 0.15, \ldots, 0.7, 0.8, \ldots,
2.0$.  We express the mean demand arrival time $\lambda$ as a function
of the offered load $\mu$ as given by (\ref{lambda}), where $\alpha$
is the mean number of links of all shortest-paths in a network begin
simulated.  We keep the mean connection holding time $\beta$ constant.
The average number of demanded slices is $\alpha{}\beta{}\gamma{} /
\lambda$, since the intensity of a demand arrival is $1 / \lambda$,
and since in an unloaded network a connection takes on average
$\alpha{}\gamma$ slices over the mean holding time of $\beta$.  The
number of slices in the network is $|E||\Omega|$, and so the offered
load $\mu = \alpha{}\beta{}\gamma{} / (\lambda{}|E||\Omega|)$, from
which (\ref{lambda}) follows.

\begin{equation}
  \lambda (\mu) = \frac{\alpha{}\beta{}\gamma{}}{\mu|E||\Omega|}
  \label{lambda}
\end{equation}

% The network utilization.

We define \emph{network utilization} as the ratio of the number of the
slices in use to the total number of slices on all edges.  We cannot
directly control the network utilization, but measure it in response
to the offered load $\mu$.

% On how we generated the random networks.

In the simulations we generated random Gabriel graphs, which can be
used to model mesh transport network topologies
\cite{10.1109/ICUMT.2013.6798402}.  Gabriel graphs can capture the
properties of mesh transport network topologies in that they are
planar, model well the length of fiber cables, the node degrees, and
the biconnected property.  We efficiently generate Gabriel graphs by
first using the Delaunay triangulation, and then removing those links
which violate the Gabriel graph properties.

% The complete reconfiguration.

The complete reconfiguration gives a chance to defragment the optical
spectrum, as the connection is completely torn down and established
anew with potentially different slices.  If the new connection uses
different slices than the previously established connection, all the
links of the new path have to be configured.  However, if the new
connection uses the same slices, then only the links not present in
the already established connection have to be configured, thus
reducing the number of the required link configurations.

% On the implementation details.

We developed and ran the software simulator under Linux within
PL-Grid, the Polish supercomputing infrastructure.  We wrote the
simulator in C++ with the Boost libraries, but without a simulation
framework, like OMNet++, as most of the functionality needed is
already provided by Boost, such as the graph manipulation with the
Boost Graph Library, or the measurement accumulation with the Boost
Accumulator.  The software is available at \cite{sdiwebsite}.

%%%%%%%%%%%%%%%%%%%%%%%%%%%%%%%%%%%%%%%%%%%%%%%%%%%%%%%%%%%%%%%%%%%%%%%%%%%

\subsection{Runs and populations}

% What's the goal.

The objective is to produce credible simulation results, which allow
us to compare the performance of the proposed reconfiguration
algorithm and the complete reconfiguration algorithm.  We carry out
simulation runs and produce credible means for statistical
populations.

% What a simulation run is.

\emph{A simulation run} simulates 100 hours of a network in operation.
There are two groups of measurements taken.  In the first group,
during every simulated hour, the \emph{measurements related to
connections} are taken when the network tries to establish or
reconfigure a connection, then the measurements are averaged and
reported for that simulated hour.  In this group, the following are
reported:
\begin{inparaenum}[(a)]
\item the probability of establishing a connection,
\item the length of an established connection,
\item the number of slices of an established connection,
\item the probability of reconfiguring a connection,
\item the number of new links of a reconfigured connection,
\item the number of reused links of a reconfigured connection,
\item the length of a reconfigured connection, and
\item the number of slices of a reconfigured connection.
\end{inparaenum}
In the second group, every simulated hour, the following
\emph{measurements related to the network} are taken:
\begin{inparaenum}[(a)]
\item the instantaneous network utilization,
\item the instantaneous number of active connections, and
\item the instantaneous amount of capacity served.
\end{inparaenum}
The measurements for every hour are averaged and reported as the
results of the simulation run.

% The population results.

\emph{A statistical population} is described by four parameters: the
reconfiguration algorithm, the routing algorithm, the spectrum
allocation policy, and the offered load $\mu$.  We have 468
populations: two reconfiguration algorithms $\times$ three routing
algorithms $\times$ three spectrum allocation policies $\times$ 26
values of the offered load $\mu$.  In a given population, all
simulation runs have the same parameters, except the seed of a random
number generator in order to generate different Gabriel graphs and
different traffic.  To get credible results for a population, we carry
out 100 simulation runs which are the population samples, and
calculate the sample means of the results reported by a simulation
run.  In total there are 46800 simulation runs (468 populations
$\times$ 100 samples).  We reckon the sample means credibly estimate
the means of their populations, since their relative standard error is
below 1\%.

\begin{table}
  \caption{Statistics of the generated Gabriel networks.}
  \label{t:netstats}
  \centering
  \begin{tabular}{|l|r|r|r|r|}
    \hline
    {\bf{}value} & {\bf{}min} & {\bf{}average} & {\bf{}max} & {\bf{}variance}\\
    \hline
    Number of links & 160 & 179.13 & 194 & 48.47\\
    Link length & 1 & 97.91 & 347 & 2695.48\\
    Node degree & 1 & 3.583 & 8 & 1.221\\
    SP length & 1 & 589.45 & 1582 & 78103.1\\
    SP hops min & 1 & 6.7667 & 22 & 10.8898\\
    \hline
  \end{tabular}
\end{table}

% On the Gabriel graphs we generate.

For the simulation runs of every population, we use the same sample of
a hundred different randomly-generated Gabriel graphs.  Each graph in
the sample has 100 nodes uniformly distributed over an area 1000 km
long and 1000 km wide, with undirected edges having 400 slices.  In
generating Gabriel graphs, the number of edges and their length cannot
be directly controlled, as it depends on the random location of nodes,
and on the candidate edges meeting the conditions of the Gabriel
graph.  Table \ref{t:netstats} reports the statistics of the generated
graphs.

%%%%%%%%%%%%%%%%%%%%%%%%%%%%%%%%%%%%%%%%%%%%%%%%%%%%%%%%%%%%%%%%%%%%%%%%%%%

\subsection{Simulation results}
\label{results}

% What results were obtained and how they are plotted.

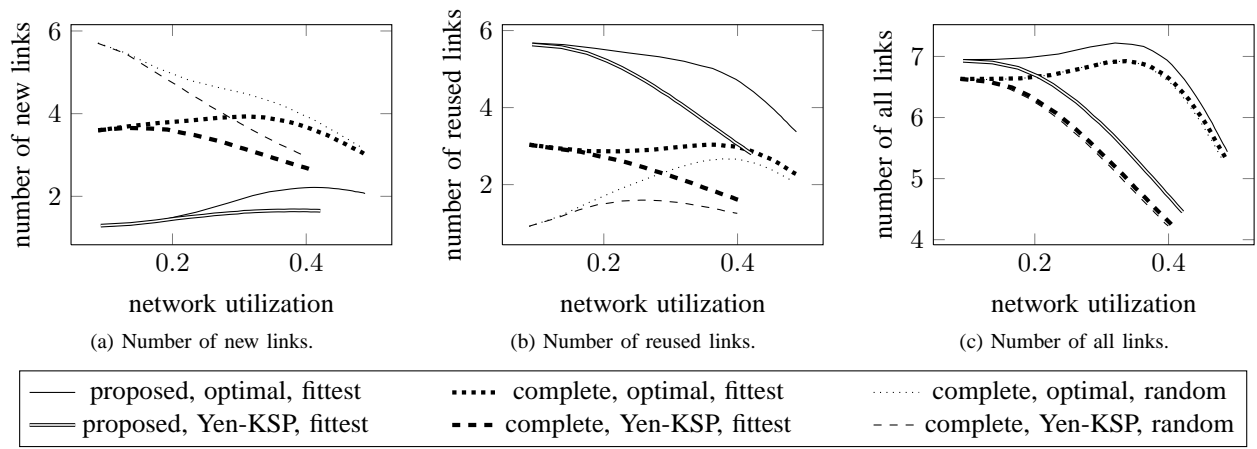
\begin{figure*}
  \centering
  \begin{tabular}{ccc}
    \subfloat[Number of new links.]{%
    \label{nonl}%
    \begin{tikzpicture}
\begin{axis}[xlabel = network utilization, ylabel style = {align = center},
ylabel = {number of new links},
legend columns = 3,
legend to name = legend,
legend style = {/tikz/every even column/.append style = {column sep = 1 cm}},
height = 4.5 cm, width = 5.8 cm,
every axis y label/.style = {at = {(ticklabel cs:0.5)}, rotate = 90, anchor = near ticklabel},
y tick label style = {/pgf/number format/fixed, /pgf/number format/1000 sep = \thinspace}]
\addplot[solid]
coordinates {
(0.0930980107075, 1.27242454142) (0.139769706066, 1.316455317) (0.187154808557, 1.4399046633) (0.23544713972, 1.6165606212) (0.28295978117, 1.8366104364) (0.32006561615, 2.0178536232) (0.34513922436, 2.1105258699) (0.36267049764, 2.1471142548) (0.37644639512, 2.1777380085) (0.38813917994, 2.1971405977) (0.39767243456, 2.2078418612) (0.40503065699, 2.2137941422) (0.41170651692, 2.2160834071) (0.42380551286, 2.2122234389) (0.43323462252, 2.204347251) (0.44131654642, 2.1892177632) (0.4484267895, 2.1808230088) (0.45447479961, 2.164773135) (0.4600625945, 2.1540988206) (0.4653046265, 2.1420783651) (0.4691643869, 2.131314623) (0.4736731003, 2.111318136) (0.4775433122, 2.107130268) (0.4812047, 2.095053979) (0.4850870407, 2.085742877) (0.4880015873, 2.068309956) 
};
\addlegendentry{proposed, optimal, fittest}
\addplot[dotted, ultra thick]
coordinates {
(0.0888911529767, 3.59588851051) (0.133383032563, 3.6925066515) (0.178273208838, 3.7726493022) (0.22411181904, 3.8307328126) (0.27113673088, 3.902890187) (0.31041925758, 3.933706757) (0.33688161698, 3.909618553) (0.35683552133, 3.853402077) (0.37131461717, 3.8130462) (0.38258526804, 3.764769391) (0.3928364426, 3.713408914) (0.40124901286, 3.668892374) (0.4087763021, 3.622579752) (0.42050535919, 3.546445718) (0.43080108754, 3.485288452) (0.43879934924, 3.425568585) (0.44633686656, 3.358503889) (0.45294031425, 3.30798784) (0.45874015308, 3.264037965) (0.46372810715, 3.219838951) (0.4683725219, 3.179638084) (0.4725115637, 3.146948443) (0.4766756751, 3.112093078) (0.4803704032, 3.088129024) (0.4840657207, 3.055535144) (0.4873111937, 3.021189387) 
};
\addlegendentry{complete, optimal, fittest}
\addplot[dotted]
coordinates {
(0.088676007178, 5.709014808) (0.132029937323, 5.4549981545) (0.17723172322, 5.1049303307) (0.22473007689, 4.810356049) (0.27072910132, 4.624737775) (0.30908676044, 4.486612386) (0.33474502533, 4.361407609) (0.35364707878, 4.262069187) (0.36764634761, 4.176405332) (0.37942848723, 4.089203878) (0.38885396266, 4.014440717) (0.397210047, 3.95173719) (0.40411554041, 3.905191598) (0.41597653719, 3.81360427) (0.42616692123, 3.714226323) (0.43423803463, 3.640122673) (0.4413556577, 3.579133704) (0.44769638793, 3.524043164) (0.45349550483, 3.465230511) (0.4581461442, 3.410795059) (0.4634474961, 3.371742842) (0.4677373279, 3.336787186) (0.471152069, 3.292408238) (0.4752820256, 3.24973596) (0.4784697947, 3.222979167) (0.4818607043, 3.192886505) 
};
\addlegendentry{complete, optimal, random}
\addplot[double]
coordinates {
(0.0927485985714, 1.29114773969) (0.13662532999, 1.330390901) (0.174990362087, 1.39649756) (0.20506157502, 1.4630260506) (0.2269836236, 1.5031796925) (0.24678004574, 1.5329452596) (0.2622578113, 1.5637913397) (0.27515215106, 1.5800268775) (0.28671804033, 1.596121987) (0.29752427011, 1.6055433595) (0.30692801804, 1.6261220603) (0.3154141886, 1.6304409408) (0.32309091117, 1.6380873135) (0.33625671555, 1.646448438) (0.3476611073, 1.653101509) (0.35799279927, 1.6568927701) (0.36688369299, 1.6548484652) (0.37514074417, 1.6611008753) (0.3822672341, 1.6618945325) (0.3897416021, 1.6588023056) (0.3954464001, 1.6612063203) (0.4016719852, 1.659596011) (0.4069610457, 1.6538327871) (0.411881415, 1.6549957046) (0.4167787224, 1.6540576) (0.4211295852, 1.6505181741) 
};
\addlegendentry{proposed, Yen-KSP, fittest}
\addplot[dashed, ultra thick]
coordinates {
(0.0887849339883, 3.59954240559) (0.130526366253, 3.6559554239) (0.165881278938, 3.6403300378) (0.19244910243, 3.6148564196) (0.21431485027, 3.536289069) (0.23238868323, 3.48414715) (0.24696684074, 3.425750135) (0.26060904441, 3.371106748) (0.2717703514, 3.327632314) (0.28139810489, 3.280812525) (0.29073166394, 3.232187568) (0.29904109993, 3.186453468) (0.30662256818, 3.153754522) (0.31983074604, 3.086513167) (0.33199964623, 3.025389724) (0.34183812008, 2.977794195) (0.35039596194, 2.930060878) (0.35949204529, 2.887026177) (0.36719964698, 2.846126841) (0.37379776405, 2.812227329) (0.3799693434, 2.781373131) (0.3864117227, 2.74898076) (0.392124378, 2.721816762) (0.3975612344, 2.698829481) (0.4024455886, 2.673740322) (0.4071459387, 2.647553386) 
};
\addlegendentry{complete, Yen-KSP, fittest}
\addplot[dashed]
coordinates {
(0.088257463115, 5.703100527) (0.129610615823, 5.4553030025) (0.16287453685, 5.1183389447) (0.190199267, 4.842193644) (0.21024023291, 4.652393023) (0.22745766208, 4.466831153) (0.24253964459, 4.326187935) (0.25516827701, 4.191669148) (0.26639385516, 4.088430919) (0.27617848544, 3.999938562) (0.28535391259, 3.910514937) (0.29323575414, 3.844139163) (0.30060286961, 3.771277136) (0.31332581781, 3.656035894) (0.32559611957, 3.550212256) (0.3349183302, 3.467761758) (0.34438477188, 3.391329763) (0.35251834799, 3.325630541) (0.3599245131, 3.260621134) (0.3667486328, 3.212377346) (0.3736212662, 3.162451529) (0.379690494, 3.118534633) (0.385078086, 3.071669786) (0.3903348024, 3.032312038) (0.3949712162, 2.997871249) (0.3996146447, 2.969971304) 
};
\addlegendentry{complete, Yen-KSP, random}
\end{axis}
\end{tikzpicture}}&%
    \subfloat[Number of reused links.]{%
    \label{nool}%
    \begin{tikzpicture}
\begin{axis}[xlabel = network utilization, ylabel style = {align = center},
ylabel = {number of reused links},
legend columns = 3,
legend to name = legend,
legend style = {/tikz/every even column/.append style = {column sep = 1 cm}},
height = 4.5 cm, width = 5.8 cm,
every axis y label/.style = {at = {(ticklabel cs:0.5)}, rotate = 90, anchor = near ticklabel},
y tick label style = {/pgf/number format/fixed, /pgf/number format/1000 sep = \thinspace}]
\addplot[solid]
coordinates {
(0.0930980107075, 5.67602279954) (0.139769706066, 5.634642219) (0.187154808557, 5.537817953) (0.23544713972, 5.422946193) (0.28295978117, 5.318965977) (0.32006561615, 5.202061727) (0.34513922436, 5.08723226) (0.36267049764, 5.016607105) (0.37644639512, 4.911669137) (0.38813917994, 4.812486704) (0.39767243456, 4.732246344) (0.40503065699, 4.64945023) (0.41170651692, 4.562448734) (0.42380551286, 4.410598952) (0.43323462252, 4.272661933) (0.44131654642, 4.156064744) (0.4484267895, 4.04204258) (0.45447479961, 3.949387996) (0.4600625945, 3.851113077) (0.4653046265, 3.768334578) (0.4691643869, 3.690380699) (0.4736731003, 3.621388027) (0.4775433122, 3.544488663) (0.4812047, 3.480320685) (0.4850870407, 3.41867863) (0.4880015873, 3.370716695) 
};
\addlegendentry{proposed, optimal, fittest}
\addplot[dotted, ultra thick]
coordinates {
(0.0888911529767, 3.03592966836) (0.133383032563, 2.9399745586) (0.178273208838, 2.86855666799) (0.22411181904, 2.8642186903) (0.27113673088, 2.9030523719) (0.31041925758, 2.9738687523) (0.33688161698, 3.0135313648) (0.35683552133, 3.0378121658) (0.37131461717, 3.0336712767) (0.38258526804, 3.0055834143) (0.3928364426, 2.9956764324) (0.40124901286, 2.965064437) (0.4087763021, 2.9348392096) (0.42050535919, 2.8726359598) (0.43080108754, 2.8049146366) (0.43879934924, 2.738291392) (0.44633686656, 2.6901121569) (0.45294031425, 2.6386321264) (0.45874015308, 2.580962532) (0.46372810715, 2.531281578) (0.4683725219, 2.475980746) (0.4725115637, 2.432735784) (0.4766756751, 2.3870696707) (0.4803704032, 2.3464943873) (0.4840657207, 2.3052379264) (0.4873111937, 2.27729881) 
};
\addlegendentry{complete, optimal, fittest}
\addplot[dotted]
coordinates {
(0.088676007178, 0.9146681379) (0.132029937323, 1.16806525586) (0.17723172322, 1.53133162631) (0.22473007689, 1.88973313823) (0.27072910132, 2.19630242163) (0.30908676044, 2.4317821907) (0.33474502533, 2.5567828168) (0.35364707878, 2.6290353989) (0.36764634761, 2.6494628117) (0.37942848723, 2.6701843785) (0.38885396266, 2.660821884) (0.397210047, 2.6619975814) (0.40411554041, 2.6465710955) (0.41597653719, 2.5961762973) (0.42616692123, 2.5555548692) (0.43423803463, 2.5050890412) (0.4413556577, 2.45777395) (0.44769638793, 2.4059523981) (0.45349550483, 2.3665577924) (0.4581461442, 2.3170538937) (0.4634474961, 2.2759333144) (0.4677373279, 2.2297167903) (0.471152069, 2.2026602818) (0.4752820256, 2.1601651632) (0.4784697947, 2.1286470022) (0.4818607043, 2.0919273874) 
};
\addlegendentry{complete, optimal, random}
\addplot[double]
coordinates {
(0.0927485985714, 5.63702340619) (0.13662532999, 5.570044103) (0.174990362087, 5.416905851) (0.20506157502, 5.210645329) (0.2269836236, 5.032349611) (0.24678004574, 4.845650485) (0.2622578113, 4.692315925) (0.27515215106, 4.555461061) (0.28671804033, 4.417860973) (0.29752427011, 4.309884369) (0.30692801804, 4.187527693) (0.3154141886, 4.095053255) (0.32309091117, 3.99969148) (0.33625671555, 3.838492972) (0.3476611073, 3.695917098) (0.35799279927, 3.568095129) (0.36688369299, 3.460169159) (0.37514074417, 3.359802738) (0.3822672341, 3.270419924) (0.3897416021, 3.182136351) (0.3954464001, 3.103210656) (0.4016719852, 3.035145813) (0.4069610457, 2.975569763) (0.411881415, 2.910078994) (0.4167787224, 2.852165288) (0.4211295852, 2.803623788) 
};
\addlegendentry{proposed, Yen-KSP, fittest}
\addplot[dashed, ultra thick]
coordinates {
(0.0887849339883, 3.03311831836) (0.130526366253, 2.93023112342) (0.165881278938, 2.8513190704) (0.19244910243, 2.7383999879) (0.21431485027, 2.6598766636) (0.23238868323, 2.5855684736) (0.24696684074, 2.5121868059) (0.26060904441, 2.4446883667) (0.2717703514, 2.3739588746) (0.28139810489, 2.3283677112) (0.29073166394, 2.2770293648) (0.29904109993, 2.2319556312) (0.30662256818, 2.1857430996) (0.31983074604, 2.1104002863) (0.33199964623, 2.0391911877) (0.34183812008, 1.9724636017) (0.35039596194, 1.9154367746) (0.35949204529, 1.8634139856) (0.36719964698, 1.818041932) (0.37379776405, 1.7724345388) (0.3799693434, 1.733834372) (0.3864117227, 1.6949527706) (0.392124378, 1.6645059566) (0.3975612344, 1.6248854597) (0.4024455886, 1.597024822) (0.4071459387, 1.5696461351) 
};
\addlegendentry{complete, Yen-KSP, fittest}
\addplot[dashed]
coordinates {
(0.088257463115, 0.92015580289) (0.129610615823, 1.13206570068) (0.16287453685, 1.34448771957) (0.190199267, 1.47639792341) (0.21024023291, 1.5264960886) (0.22745766208, 1.5748924571) (0.24253964459, 1.58626563426) (0.25516827701, 1.596984591) (0.26639385516, 1.5967618005) (0.27617848544, 1.58850169476) (0.28535391259, 1.58124426875) (0.29323575414, 1.56300091052) (0.30060286961, 1.553748608) (0.31332581781, 1.5340096433) (0.32559611957, 1.50090581903) (0.3349183302, 1.4718875839) (0.34438477188, 1.4437344879) (0.35251834799, 1.4188644829) (0.3599245131, 1.3989944912) (0.3667486328, 1.3723420107) (0.3736212662, 1.3494534864) (0.379690494, 1.3199593967) (0.385078086, 1.3106791915) (0.3903348024, 1.2873536159) (0.3949712162, 1.2671870023) (0.3996146447, 1.2544748735) 
};
\addlegendentry{complete, Yen-KSP, random}
\end{axis}
\end{tikzpicture}}&%
    \subfloat[Number of all links.]{%
    \label{noal}%
    \begin{tikzpicture}
\begin{axis}[xlabel = network utilization, ylabel style = {align = center},
ylabel = {number of all links},
legend columns = 3,
legend to name = legend,
legend style = {/tikz/every even column/.append style = {column sep = 1 cm}},
height = 4.5 cm, width = 5.8 cm,
every axis y label/.style = {at = {(ticklabel cs:0.5)}, rotate = 90, anchor = near ticklabel},
y tick label style = {/pgf/number format/fixed, /pgf/number format/1000 sep = \thinspace}]
\addplot[solid]
coordinates {
(0.0930980107075, 6.94844722485) (0.139769706066, 6.951097469) (0.187154808557, 6.977722671) (0.23544713972, 7.039506855) (0.28295978117, 7.155576411) (0.32006561615, 7.219915363) (0.34513922436, 7.197758122) (0.36267049764, 7.163721295) (0.37644639512, 7.089407171) (0.38813917994, 7.009627218) (0.39767243456, 6.940088182) (0.40503065699, 6.863244399) (0.41170651692, 6.77853218) (0.42380551286, 6.622822328) (0.43323462252, 6.477009169) (0.44131654642, 6.345282522) (0.4484267895, 6.22286555) (0.45447479961, 6.114161151) (0.4600625945, 6.005211968) (0.4653046265, 5.910412886) (0.4691643869, 5.821695357) (0.4736731003, 5.732706134) (0.4775433122, 5.651618947) (0.4812047, 5.57537467) (0.4850870407, 5.504421546) (0.4880015873, 5.439026767) 
};
\addlegendentry{proposed, optimal, fittest}
\addplot[dotted, ultra thick]
coordinates {
(0.0888911529767, 6.63181816761) (0.133383032563, 6.632481205) (0.178273208838, 6.641206035) (0.22411181904, 6.694951482) (0.27113673088, 6.805942584) (0.31041925758, 6.907575476) (0.33688161698, 6.923150009) (0.35683552133, 6.891214249) (0.37131461717, 6.846717475) (0.38258526804, 6.770352871) (0.3928364426, 6.709085373) (0.40124901286, 6.63395677) (0.4087763021, 6.557418894) (0.42050535919, 6.419081642) (0.43080108754, 6.290203048) (0.43879934924, 6.163860032) (0.44633686656, 6.048616049) (0.45294031425, 5.946620056) (0.45874015308, 5.845000561) (0.46372810715, 5.751120448) (0.4683725219, 5.655618854) (0.4725115637, 5.579684206) (0.4766756751, 5.49916278) (0.4803704032, 5.434623482) (0.4840657207, 5.360773058) (0.4873111937, 5.298488157) 
};
\addlegendentry{complete, optimal, fittest}
\addplot[dotted]
coordinates {
(0.088676007178, 6.623682883) (0.132029937323, 6.623063361) (0.17723172322, 6.636262088) (0.22473007689, 6.700089216) (0.27072910132, 6.821040246) (0.30908676044, 6.91839462) (0.33474502533, 6.918190377) (0.35364707878, 6.891104576) (0.36764634761, 6.825868067) (0.37942848723, 6.759388283) (0.38885396266, 6.675262613) (0.397210047, 6.613734786) (0.40411554041, 6.551762669) (0.41597653719, 6.40978059) (0.42616692123, 6.269781216) (0.43423803463, 6.145211777) (0.4413556577, 6.036907653) (0.44769638793, 5.929995543) (0.45349550483, 5.831788293) (0.4581461442, 5.727848892) (0.4634474961, 5.647676138) (0.4677373279, 5.566504054) (0.471152069, 5.495068577) (0.4752820256, 5.409901063) (0.4784697947, 5.35162616) (0.4818607043, 5.284813946) 
};
\addlegendentry{complete, optimal, random}
\addplot[double]
coordinates {
(0.0927485985714, 6.92817111662) (0.13662532999, 6.900434979) (0.174990362087, 6.813403397) (0.20506157502, 6.673671384) (0.2269836236, 6.535529297) (0.24678004574, 6.3785957) (0.2622578113, 6.256107327) (0.27515215106, 6.135487955) (0.28671804033, 6.013982951) (0.29752427011, 5.915427666) (0.30692801804, 5.813649749) (0.3154141886, 5.725494179) (0.32309091117, 5.637778846) (0.33625671555, 5.484941421) (0.3476611073, 5.349018605) (0.35799279927, 5.224987869) (0.36688369299, 5.115017677) (0.37514074417, 5.020903656) (0.3822672341, 4.932314398) (0.3897416021, 4.840938709) (0.3954464001, 4.764417023) (0.4016719852, 4.694741768) (0.4069610457, 4.629402561) (0.411881415, 4.565074668) (0.4167787224, 4.506222818) (0.4211295852, 4.454142042) 
};
\addlegendentry{proposed, Yen-KSP, fittest}
\addplot[dashed, ultra thick]
coordinates {
(0.0887849339883, 6.63266073104) (0.130526366253, 6.586186597) (0.165881278938, 6.491649063) (0.19244910243, 6.353256457) (0.21431485027, 6.196165704) (0.23238868323, 6.069715637) (0.24696684074, 5.937937008) (0.26060904441, 5.815795094) (0.2717703514, 5.701591185) (0.28139810489, 5.609180255) (0.29073166394, 5.509216943) (0.29904109993, 5.418409075) (0.30662256818, 5.339497653) (0.31983074604, 5.196913512) (0.33199964623, 5.064580826) (0.34183812008, 4.950257844) (0.35039596194, 4.845497678) (0.35949204529, 4.750440092) (0.36719964698, 4.664168878) (0.37379776405, 4.584661888) (0.3799693434, 4.515207493) (0.3864117227, 4.443933538) (0.392124378, 4.386322648) (0.3975612344, 4.323714917) (0.4024455886, 4.270765117) (0.4071459387, 4.217199542) 
};
\addlegendentry{complete, Yen-KSP, fittest}
\addplot[dashed]
coordinates {
(0.088257463115, 6.623256312) (0.129610615823, 6.58736869) (0.16287453685, 6.462826674) (0.190199267, 6.318591528) (0.21024023291, 6.178889013) (0.22745766208, 6.04172359) (0.24253964459, 5.912453647) (0.25516827701, 5.788653741) (0.26639385516, 5.685192711) (0.27617848544, 5.588440273) (0.28535391259, 5.491759103) (0.29323575414, 5.407140087) (0.30060286961, 5.325025764) (0.31332581781, 5.190045574) (0.32559611957, 5.051118141) (0.3349183302, 4.939649288) (0.34438477188, 4.835064178) (0.35251834799, 4.744495002) (0.3599245131, 4.659615643) (0.3667486328, 4.584719339) (0.3736212662, 4.511904972) (0.379690494, 4.438494038) (0.385078086, 4.382348921) (0.3903348024, 4.319665584) (0.3949712162, 4.265058156) (0.3996146447, 4.22444623) 
};
\addlegendentry{complete, Yen-KSP, random}
\end{axis}
\end{tikzpicture}}\\[15pt]
  \end{tabular}
  \ref{legend}\\[5pt]
  \caption{The number of new, reused, and all links of a reconfigured
  connection.}
  \label{first}
\end{figure*}

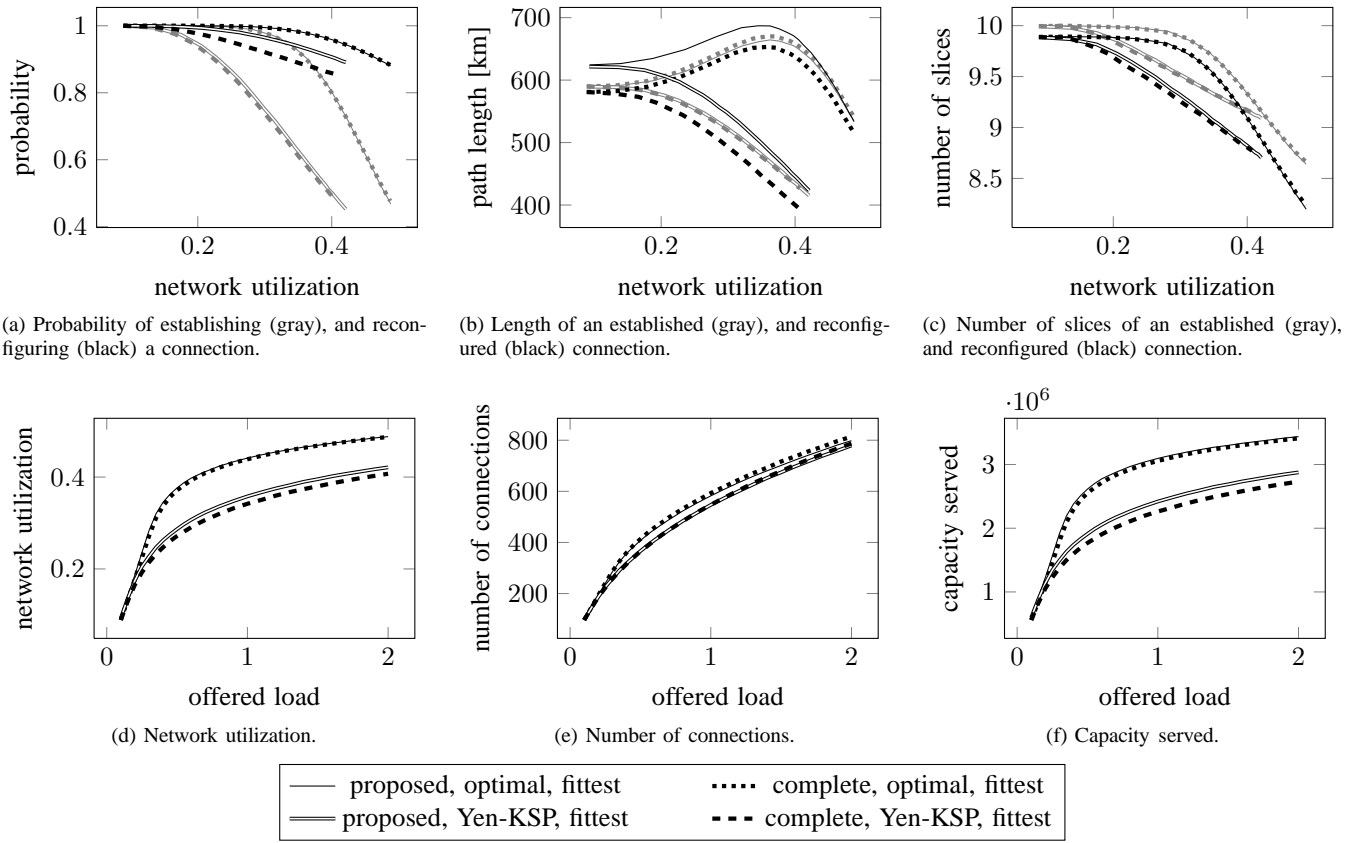
\begin{figure*}
  \centering
  \begin{tabular}{ccc}

    \subfloat[Probability of establishing (gray), and reconfiguring (black) a connection.]{%
    \label{probs}%
    \begin{tikzpicture}
\begin{axis}[xlabel = network utilization, ylabel style = {align = center},
ylabel = {probability},
legend columns = 2,
legend to name = legend,
legend style = {/tikz/every even column/.append style = {column sep = 1 cm}},
height = 4.5 cm, width = 5.8 cm,
every axis y label/.style = {at = {(ticklabel cs:0.5)}, rotate = 90, anchor = near ticklabel},
y tick label style = {/pgf/number format/fixed, /pgf/number format/1000 sep = \thinspace}]
\addplot[solid, gray]
coordinates {
(0.0930980107075, 1.0) (0.139769706066, 0.9999952381) (0.187154808557, 0.9999572779) (0.23544713972, 0.9984411575) (0.28295978117, 0.9885935045) (0.32006561615, 0.9660169042) (0.34513922436, 0.9334970573) (0.36267049764, 0.9028044946) (0.37644639512, 0.8684239257) (0.38813917994, 0.838680628) (0.39767243456, 0.8113210757) (0.40503065699, 0.7851225131) (0.41170651692, 0.7621569186) (0.42380551286, 0.7188047493) (0.43323462252, 0.6814144153) (0.44131654642, 0.6501750315) (0.4484267895, 0.6223004307) (0.45447479961, 0.598600177) (0.4600625945, 0.5762011425) (0.4653046265, 0.5556231617) (0.4691643869, 0.5388817236) (0.4736731003, 0.5214590727) (0.4775433122, 0.5069638459) (0.4812047, 0.4935915275) (0.4850870407, 0.4800488671) (0.4880015873, 0.4680246783) 
};
\addlegendentry{proposed, optimal, fittest}
\addplot[dotted, ultra thick, gray]
coordinates {
(0.0888911529767, 1.0) (0.133383032563, 0.9999833333) (0.178273208838, 0.9999083341) (0.22411181904, 0.9988715982) (0.27113673088, 0.9925182321) (0.31041925758, 0.9722298068) (0.33688161698, 0.9440546587) (0.35683552133, 0.9109294911) (0.37131461717, 0.8800340788) (0.38258526804, 0.8505618439) (0.3928364426, 0.8213117472) (0.40124901286, 0.7958631237) (0.4087763021, 0.7726043308) (0.42050535919, 0.7302621745) (0.43080108754, 0.6924879757) (0.43879934924, 0.6610003536) (0.44633686656, 0.632064216) (0.45294031425, 0.606456973) (0.45874015308, 0.5846424732) (0.46372810715, 0.5656876612) (0.4683725219, 0.5467337541) (0.4725115637, 0.5301597792) (0.4766756751, 0.5155182343) (0.4803704032, 0.5014267893) (0.4840657207, 0.4890036204) (0.4873111937, 0.4769826384) 
};
\addlegendentry{complete, optimal, fittest}
\addplot[double, gray]
coordinates {
(0.0927485985714, 0.9998221311) (0.13662532999, 0.9938432637) (0.174990362087, 0.9731105964) (0.20506157502, 0.9400545695) (0.2269836236, 0.9061113705) (0.24678004574, 0.8705265568) (0.2622578113, 0.8401712011) (0.27515215106, 0.8109957109) (0.28671804033, 0.7834063634) (0.29752427011, 0.7567808641) (0.30692801804, 0.7351783566) (0.3154141886, 0.7153530189) (0.32309091117, 0.6955411655) (0.33625671555, 0.6610087523) (0.3476611073, 0.6314067476) (0.35799279927, 0.6058107416) (0.36688369299, 0.5828038207) (0.37514074417, 0.5619542729) (0.3822672341, 0.5442668266) (0.3897416021, 0.5270058528) (0.3954464001, 0.5111878072) (0.4016719852, 0.4981406863) (0.4069610457, 0.4853637546) (0.411881415, 0.4733072071) (0.4167787224, 0.4622546775) (0.4211295852, 0.452320054) 
};
\addlegendentry{proposed, Yen-KSP, fittest}
\addplot[dashed, ultra thick, gray]
coordinates {
(0.0887849339883, 0.9998641357) (0.130526366253, 0.9948634417) (0.165881278938, 0.9753909002) (0.19244910243, 0.9484011746) (0.21431485027, 0.917140894) (0.23238868323, 0.8866741904) (0.24696684074, 0.8579734789) (0.26060904441, 0.8300515662) (0.2717703514, 0.8049032782) (0.28139810489, 0.7820353537) (0.29073166394, 0.7609349328) (0.29904109993, 0.7411256462) (0.30662256818, 0.7219559821) (0.31983074604, 0.6887712176) (0.33199964623, 0.6597902187) (0.34183812008, 0.6338640619) (0.35039596194, 0.6106569427) (0.35949204529, 0.5894452016) (0.36719964698, 0.5701724167) (0.37379776405, 0.5550752568) (0.3799693434, 0.5377533295) (0.3864117227, 0.5238362182) (0.392124378, 0.5111518024) (0.3975612344, 0.4981588741) (0.4024455886, 0.487236272) (0.4071459387, 0.4766597615) 
};
\addlegendentry{complete, Yen-KSP, fittest}
\addplot[solid]
coordinates {
(0.0930980107075, 1.0) (0.139769706066, 0.999995) (0.187154808557, 0.9999391018) (0.23544713972, 0.9994790724) (0.28295978117, 0.9970465576) (0.32006561615, 0.9915778855) (0.34513922436, 0.9848744032) (0.36267049764, 0.9780786967) (0.37644639512, 0.9714049027) (0.38813917994, 0.96568219) (0.39767243456, 0.9602159944) (0.40503065699, 0.9556861141) (0.41170651692, 0.950519862) (0.42380551286, 0.9414777388) (0.43323462252, 0.9340615618) (0.44131654642, 0.926862278) (0.4484267895, 0.9203894224) (0.45447479961, 0.9145345902) (0.4600625945, 0.9098564432) (0.4653046265, 0.9044329084) (0.4691643869, 0.8999365084) (0.4736731003, 0.8953490822) (0.4775433122, 0.8916000367) (0.4812047, 0.8878377558) (0.4850870407, 0.8838568043) (0.4880015873, 0.8808320614) 
};
\addlegendentry{proposed, optimal, fittest}
\addplot[dotted, ultra thick]
coordinates {
(0.0888911529767, 1.0) (0.133383032563, 1.0) (0.178273208838, 0.9999809074) (0.22411181904, 0.9997280092) (0.27113673088, 0.9979133507) (0.31041925758, 0.9932649643) (0.33688161698, 0.9866975719) (0.35683552133, 0.9798996118) (0.37131461717, 0.9737890018) (0.38258526804, 0.9680254482) (0.3928364426, 0.9621024502) (0.40124901286, 0.9578003124) (0.4087763021, 0.9519348765) (0.42050535919, 0.9442177063) (0.43080108754, 0.9358943617) (0.43879934924, 0.9292503168) (0.44633686656, 0.9228337838) (0.45294031425, 0.9170867654) (0.45874015308, 0.9116935466) (0.46372810715, 0.9065766501) (0.4683725219, 0.9021073383) (0.4725115637, 0.8980671417) (0.4766756751, 0.8940346436) (0.4803704032, 0.8900797258) (0.4840657207, 0.8864712975) (0.4873111937, 0.8827381749) 
};
\addlegendentry{complete, optimal, fittest}
\addplot[double]
coordinates {
(0.0927485985714, 0.9999857143) (0.13662532999, 0.9995043211) (0.174990362087, 0.9971845476) (0.20506157502, 0.9935916132) (0.2269836236, 0.9891501345) (0.24678004574, 0.9841524413) (0.2622578113, 0.9800494542) (0.27515215106, 0.9748696361) (0.28671804033, 0.9706493993) (0.29752427011, 0.9661389986) (0.30692801804, 0.9615820417) (0.3154141886, 0.9575100317) (0.32309091117, 0.953694794) (0.33625671555, 0.9470350117) (0.3476611073, 0.9405508808) (0.35799279927, 0.9339631007) (0.36688369299, 0.9286411451) (0.37514074417, 0.9232171115) (0.3822672341, 0.918526689) (0.3897416021, 0.9134220231) (0.3954464001, 0.9097977806) (0.4016719852, 0.9055871231) (0.4069610457, 0.9017829316) (0.411881415, 0.8976033428) (0.4167787224, 0.8936183732) (0.4211295852, 0.891238352) 
};
\addlegendentry{proposed, Yen-KSP, fittest}
\addplot[dashed, ultra thick]
coordinates {
(0.0887849339883, 0.9999156377) (0.130526366253, 0.9976560794) (0.165881278938, 0.9899298889) (0.19244910243, 0.9796768986) (0.21431485027, 0.970277391) (0.23238868323, 0.9613415043) (0.24696684074, 0.9531163198) (0.26060904441, 0.9454204535) (0.2717703514, 0.9384741786) (0.28139810489, 0.9324027123) (0.29073166394, 0.9272452015) (0.29904109993, 0.922000541) (0.30662256818, 0.9166628556) (0.31983074604, 0.9091348884) (0.33199964623, 0.9014026535) (0.34183812008, 0.8949721536) (0.35039596194, 0.8890610578) (0.35949204529, 0.8835801114) (0.36719964698, 0.8791071185) (0.37379776405, 0.8754398954) (0.3799693434, 0.8707245966) (0.3864117227, 0.8665706673) (0.392124378, 0.8625481416) (0.3975612344, 0.8598822696) (0.4024455886, 0.8572085666) (0.4071459387, 0.8535460312) 
};
\addlegendentry{complete, Yen-KSP, fittest}
\end{axis}
\end{tikzpicture}}&%
    \subfloat[Length of an established (gray), and reconfigured (black) connection.]{%
    \label{lengths}%
    \begin{tikzpicture}
\begin{axis}[xlabel = network utilization, ylabel style = {align = center},
ylabel = {path length [km]},
legend columns = 2,
legend to name = legend,
legend style = {/tikz/every even column/.append style = {column sep = 1 cm}},
height = 4.5 cm, width = 5.8 cm,
every axis y label/.style = {at = {(ticklabel cs:0.5)}, rotate = 90, anchor = near ticklabel},
y tick label style = {/pgf/number format/fixed, /pgf/number format/1000 sep = \thinspace}]
\addplot[solid, gray]
coordinates {
(0.0930980107075, 589.771316901) (0.139769706066, 591.6310661) (0.187154808557, 598.7629401) (0.23544713972, 613.5338658) (0.28295978117, 636.9659637) (0.32006561615, 654.4119167) (0.34513922436, 661.3550767) (0.36267049764, 664.3382095) (0.37644639512, 662.4046227) (0.38813917994, 658.4647407) (0.39767243456, 655.2015003) (0.40503065699, 649.813919) (0.41170651692, 644.0582902) (0.42380551286, 633.3738516) (0.43323462252, 621.9837436) (0.44131654642, 611.8872761) (0.4484267895, 601.8446837) (0.45447479961, 592.6053994) (0.4600625945, 583.6424968) (0.4653046265, 575.4225831) (0.4691643869, 567.7974306) (0.4736731003, 559.8994586) (0.4775433122, 552.8584374) (0.4812047, 546.3397549) (0.4850870407, 539.5228255) (0.4880015873, 534.3941773) 
};
\addlegendentry{proposed, optimal, fittest}
\addplot[dotted, ultra thick, gray]
coordinates {
(0.0888911529767, 589.773497501) (0.133383032563, 591.6099878) (0.178273208838, 598.3746686) (0.22411181904, 612.6163786) (0.27113673088, 635.2709524) (0.31041925758, 655.7285716) (0.33688161698, 666.3246746) (0.35683552133, 669.8334371) (0.37131461717, 670.1087208) (0.38258526804, 665.8920308) (0.3928364426, 662.4900812) (0.40124901286, 657.6162438) (0.4087763021, 652.0297593) (0.42050535919, 640.9429962) (0.43080108754, 630.8025765) (0.43879934924, 619.9351432) (0.44633686656, 609.6528365) (0.45294031425, 601.2390782) (0.45874015308, 592.0384287) (0.46372810715, 583.3271972) (0.4683725219, 575.222797) (0.4725115637, 567.9306581) (0.4766756751, 560.4986198) (0.4803704032, 554.709429) (0.4840657207, 547.6255275) (0.4873111937, 542.4258973) 
};
\addlegendentry{complete, optimal, fittest}
\addplot[double, gray]
coordinates {
(0.0927485985714, 589.456156001) (0.13662532999, 588.624167) (0.174990362087, 584.0531635) (0.20506157502, 575.2500206) (0.2269836236, 566.9656164) (0.24678004574, 556.2739833) (0.2622578113, 548.1231538) (0.27515215106, 539.8404087) (0.28671804033, 531.4187676) (0.29752427011, 524.289168) (0.30692801804, 516.6655408) (0.3154141886, 510.4293046) (0.32309091117, 503.9173164) (0.33625671555, 492.7702444) (0.3476611073, 482.9836888) (0.35799279927, 473.5473027) (0.36688369299, 465.4153058) (0.37514074417, 458.3318823) (0.3822672341, 451.9154841) (0.3897416021, 444.4310386) (0.3954464001, 438.7178296) (0.4016719852, 433.5763308) (0.4069610457, 428.5524148) (0.411881415, 423.7398795) (0.4167787224, 419.1214726) (0.4211295852, 415.1659148) 
};
\addlegendentry{proposed, Yen-KSP, fittest}
\addplot[dashed, ultra thick, gray]
coordinates {
(0.0887849339883, 589.866427701) (0.130526366253, 587.9962721) (0.165881278938, 584.9091149) (0.19244910243, 578.2367073) (0.21431485027, 569.5831697) (0.23238868323, 562.1142266) (0.24696684074, 554.1974966) (0.26060904441, 546.2610486) (0.2717703514, 538.7888929) (0.28139810489, 532.7177086) (0.29073166394, 525.8842218) (0.29904109993, 519.5133188) (0.30662256818, 513.8315764) (0.31983074604, 503.6901096) (0.33199964623, 493.1920825) (0.34183812008, 485.0018843) (0.35039596194, 476.9517201) (0.35949204529, 469.5284705) (0.36719964698, 462.8817687) (0.37379776405, 456.6764458) (0.3799693434, 450.8304443) (0.3864117227, 444.9356505) (0.392124378, 440.6355076) (0.3975612344, 435.4660445) (0.4024455886, 430.8869184) (0.4071459387, 426.5781211) 
};
\addlegendentry{complete, Yen-KSP, fittest}
\addplot[solid]
coordinates {
(0.0930980107075, 624.331038201) (0.139769706066, 626.4881787) (0.187154808557, 634.5487967) (0.23544713972, 648.8175979) (0.28295978117, 669.7644133) (0.32006561615, 683.9073281) (0.34513922436, 686.965016) (0.36267049764, 686.7096037) (0.37644639512, 681.9933125) (0.38813917994, 676.2829013) (0.39767243456, 671.0906644) (0.40503065699, 664.804995) (0.41170651692, 657.1192982) (0.42380551286, 644.0635669) (0.43323462252, 630.9619795) (0.44131654642, 618.797033) (0.4484267895, 607.5731789) (0.45447479961, 597.5483409) (0.4600625945, 587.3655673) (0.4653046265, 578.6361593) (0.4691643869, 569.8420597) (0.4736731003, 561.3408976) (0.4775433122, 553.8820674) (0.4812047, 546.4439075) (0.4850870407, 539.7585159) (0.4880015873, 533.5938306) 
};
\addlegendentry{proposed, optimal, fittest}
\addplot[dotted, ultra thick]
coordinates {
(0.0888911529767, 580.891412785) (0.133383032563, 582.3565433) (0.178273208838, 589.1082874) (0.22411181904, 603.0714768) (0.27113673088, 624.5264779) (0.31041925758, 643.6258304) (0.33688161698, 651.5559448) (0.35683552133, 653.0101906) (0.37131461717, 651.8137883) (0.38258526804, 646.4290288) (0.3928364426, 642.1063304) (0.40124901286, 636.4912924) (0.4087763021, 630.1957347) (0.42050535919, 618.5423137) (0.43080108754, 607.5907176) (0.43879934924, 596.2523586) (0.44633686656, 585.7013265) (0.45294031425, 576.5615562) (0.45874015308, 567.3766403) (0.46372810715, 558.6425763) (0.4683725219, 549.7692326) (0.4725115637, 542.6635724) (0.4766756751, 535.1438577) (0.4803704032, 529.1388068) (0.4840657207, 522.2371993) (0.4873111937, 516.3374443) 
};
\addlegendentry{complete, optimal, fittest}
\addplot[double]
coordinates {
(0.0927485985714, 621.565646627) (0.13662532999, 620.5317716) (0.174990362087, 615.997282) (0.20506157502, 606.4910789) (0.2269836236, 597.3325641) (0.24678004574, 585.305023) (0.2622578113, 576.0963139) (0.27515215106, 566.3573186) (0.28671804033, 556.5940523) (0.29752427011, 548.7465132) (0.30692801804, 540.4019626) (0.3154141886, 533.1247473) (0.32309091117, 525.6886301) (0.33625671555, 512.7117042) (0.3476611073, 501.4228764) (0.35799279927, 490.5782855) (0.36688369299, 481.2510706) (0.37514074417, 472.9974279) (0.3822672341, 465.5281827) (0.3897416021, 457.3106885) (0.3954464001, 450.5158319) (0.4016719852, 444.4126301) (0.4069610457, 438.6195726) (0.411881415, 432.9806535) (0.4167787224, 427.628723) (0.4211295852, 423.1473322) 
};
\addlegendentry{proposed, Yen-KSP, fittest}
\addplot[dashed, ultra thick]
coordinates {
(0.0887849339883, 581.006568885) (0.130526366253, 578.2847603) (0.165881278938, 573.2038568) (0.19244910243, 564.0147512) (0.21431485027, 552.9959797) (0.23238868323, 543.9731247) (0.24696684074, 534.2179725) (0.26060904441, 524.6472982) (0.2717703514, 515.7872806) (0.28139810489, 508.6105277) (0.29073166394, 500.7680129) (0.29904109993, 493.5333797) (0.30662256818, 487.1184751) (0.31983074604, 475.9077172) (0.33199964623, 464.5644713) (0.34183812008, 455.0771865) (0.35039596194, 446.381314) (0.35949204529, 438.5594646) (0.36719964698, 431.4585685) (0.37379776405, 424.640277) (0.3799693434, 418.7751338) (0.3864117227, 412.7071014) (0.392124378, 407.7439857) (0.3975612344, 402.3340132) (0.4024455886, 397.8682616) (0.4071459387, 393.3580492) 
};
\addlegendentry{complete, Yen-KSP, fittest}
\end{axis}
\end{tikzpicture}}&%
    \subfloat[Number of slices of an established (gray), and reconfigured (black) connection.]{%
    \label{slices}%
    \begin{tikzpicture}
\begin{axis}[xlabel = network utilization, ylabel style = {align = center},
ylabel = {number of slices},
legend columns = 2,
legend to name = legend,
legend style = {/tikz/every even column/.append style = {column sep = 1 cm}},
height = 4.5 cm, width = 5.8 cm,
every axis y label/.style = {at = {(ticklabel cs:0.5)}, rotate = 90, anchor = near ticklabel},
y tick label style = {/pgf/number format/fixed, /pgf/number format/1000 sep = \thinspace}]
\addplot[solid, gray]
coordinates {
(0.0930980107075, 9.99539475529) (0.139769706066, 9.99789207) (0.187154808557, 9.990303778) (0.23544713972, 9.980910088) (0.28295978117, 9.942216029) (0.32006561615, 9.836505116) (0.34513922436, 9.72720933) (0.36267049764, 9.616711721) (0.37644639512, 9.518154598) (0.38813917994, 9.436275535) (0.39767243456, 9.352622857) (0.40503065699, 9.285119466) (0.41170651692, 9.235734663) (0.42380551286, 9.131937246) (0.43323462252, 9.05137662) (0.44131654642, 8.983216448) (0.4484267895, 8.932516898) (0.45447479961, 8.877248952) (0.4600625945, 8.831983913) (0.4653046265, 8.796255083) (0.4691643869, 8.764710443) (0.4736731003, 8.737092122) (0.4775433122, 8.708921125) (0.4812047, 8.687036846) (0.4850870407, 8.663088759) (0.4880015873, 8.638388535) 
};
\addlegendentry{proposed, optimal, fittest}
\addplot[dotted, ultra thick, gray]
coordinates {
(0.0888911529767, 9.99539475529) (0.133383032563, 9.99842638) (0.178273208838, 9.989876789) (0.22411181904, 9.981151425) (0.27113673088, 9.947584145) (0.31041925758, 9.875973922) (0.33688161698, 9.76176815) (0.35683552133, 9.653566389) (0.37131461717, 9.550582654) (0.38258526804, 9.466359525) (0.3928364426, 9.387814075) (0.40124901286, 9.322343509) (0.4087763021, 9.265500216) (0.42050535919, 9.164453895) (0.43080108754, 9.080133439) (0.43879934924, 9.008296203) (0.44633686656, 8.952098548) (0.45294031425, 8.899393304) (0.45874015308, 8.857365324) (0.46372810715, 8.825798426) (0.4683725219, 8.790258852) (0.4725115637, 8.754694852) (0.4766756751, 8.7326927) (0.4803704032, 8.711071584) (0.4840657207, 8.684309981) (0.4873111937, 8.668773289) 
};
\addlegendentry{complete, optimal, fittest}
\addplot[double, gray]
coordinates {
(0.0927485985714, 9.99154411429) (0.13662532999, 9.977327699) (0.174990362087, 9.928231383) (0.20506157502, 9.855435583) (0.2269836236, 9.786102575) (0.24678004574, 9.71353558) (0.2622578113, 9.66392339) (0.27515215106, 9.613563137) (0.28671804033, 9.563711382) (0.29752427011, 9.528935146) (0.30692801804, 9.494197742) (0.3154141886, 9.464362948) (0.32309091117, 9.428925444) (0.33625671555, 9.383865352) (0.3476611073, 9.337041087) (0.35799279927, 9.302869099) (0.36688369299, 9.269367563) (0.37514074417, 9.243281266) (0.3822672341, 9.214379799) (0.3897416021, 9.197019317) (0.3954464001, 9.174052406) (0.4016719852, 9.158262627) (0.4069610457, 9.140019117) (0.411881415, 9.120804646) (0.4167787224, 9.112534694) (0.4211295852, 9.090205114) 
};
\addlegendentry{proposed, Yen-KSP, fittest}
\addplot[dashed, ultra thick, gray]
coordinates {
(0.0887849339883, 9.99225950229) (0.130526366253, 9.998565525) (0.165881278938, 9.93582482) (0.19244910243, 9.873826729) (0.21431485027, 9.797892793) (0.23238868323, 9.735948203) (0.24696684074, 9.697605722) (0.26060904441, 9.647572495) (0.2717703514, 9.605585782) (0.28139810489, 9.578211291) (0.29073166394, 9.534306376) (0.29904109993, 9.512223338) (0.30662256818, 9.480568145) (0.31983074604, 9.433394034) (0.33199964623, 9.396423339) (0.34183812008, 9.36027416) (0.35039596194, 9.32618383) (0.35949204529, 9.30039437) (0.36719964698, 9.273659382) (0.37379776405, 9.251811864) (0.3799693434, 9.229960684) (0.3864117227, 9.209891017) (0.392124378, 9.19239302) (0.3975612344, 9.178782619) (0.4024455886, 9.161965108) (0.4071459387, 9.149568448) 
};
\addlegendentry{complete, Yen-KSP, fittest}
\addplot[solid]
coordinates {
(0.0930980107075, 9.88900203327) (0.139769706066, 9.891791161) (0.187154808557, 9.88739345) (0.23544713972, 9.87240972) (0.28295978117, 9.821613562) (0.32006561615, 9.691334754) (0.34513922436, 9.562720991) (0.36267049764, 9.431217544) (0.37644639512, 9.317732798) (0.38813917994, 9.215524376) (0.39767243456, 9.121534033) (0.40503065699, 9.039726829) (0.41170651692, 8.975713461) (0.42380551286, 8.849913256) (0.43323462252, 8.752497002) (0.44131654642, 8.667167291) (0.4484267895, 8.600052052) (0.45447479961, 8.530222024) (0.4600625945, 8.468411324) (0.4653046265, 8.421178992) (0.4691643869, 8.378179468) (0.4736731003, 8.336205553) (0.4775433122, 8.299880462) (0.4812047, 8.265988676) (0.4850870407, 8.227584776) (0.4880015873, 8.198090323) 
};
\addlegendentry{proposed, optimal, fittest}
\addplot[dotted, ultra thick]
coordinates {
(0.0888911529767, 9.88900203327) (0.133383032563, 9.892188651) (0.178273208838, 9.886848556) (0.22411181904, 9.874942515) (0.27113673088, 9.834246046) (0.31041925758, 9.743535894) (0.33688161698, 9.605901945) (0.35683552133, 9.473880313) (0.37131461717, 9.35793292) (0.38258526804, 9.251386733) (0.3928364426, 9.157562649) (0.40124901286, 9.083410027) (0.4087763021, 9.01082094) (0.42050535919, 8.891085714) (0.43080108754, 8.788840573) (0.43879934924, 8.697741571) (0.44633686656, 8.622696848) (0.45294031425, 8.557576719) (0.45874015308, 8.501881644) (0.46372810715, 8.460244298) (0.4683725219, 8.411329868) (0.4725115637, 8.3627315) (0.4766756751, 8.333264212) (0.4803704032, 8.301315912) (0.4840657207, 8.264705372) (0.4873111937, 8.234882731) 
};
\addlegendentry{complete, optimal, fittest}
\addplot[double]
coordinates {
(0.0927485985714, 9.88416613727) (0.13662532999, 9.866442201) (0.174990362087, 9.815641201) (0.20506157502, 9.726175091) (0.2269836236, 9.644705891) (0.24678004574, 9.562085971) (0.2622578113, 9.498322756) (0.27515215106, 9.434674714) (0.28671804033, 9.377016378) (0.29752427011, 9.332078257) (0.30692801804, 9.286905458) (0.3154141886, 9.247026253) (0.32309091117, 9.206116062) (0.33625671555, 9.14210662) (0.3476611073, 9.076153091) (0.35799279927, 9.031280244) (0.36688369299, 8.982183444) (0.37514074417, 8.942146987) (0.3822672341, 8.903685585) (0.3897416021, 8.874387005) (0.3954464001, 8.843419619) (0.4016719852, 8.81332169) (0.4069610457, 8.782095447) (0.411881415, 8.754327346) (0.4167787224, 8.735476676) (0.4211295852, 8.703220313) 
};
\addlegendentry{proposed, Yen-KSP, fittest}
\addplot[dashed, ultra thick]
coordinates {
(0.0887849339883, 9.88580049127) (0.130526366253, 9.885660001) (0.165881278938, 9.802002684) (0.19244910243, 9.723637113) (0.21431485027, 9.625207079) (0.23238868323, 9.553459735) (0.24696684074, 9.498758529) (0.26060904441, 9.437414275) (0.2717703514, 9.385673079) (0.28139810489, 9.348686101) (0.29073166394, 9.299792451) (0.29904109993, 9.262370155) (0.30662256818, 9.226851779) (0.31983074604, 9.170320295) (0.33199964623, 9.119889608) (0.34183812008, 9.07142749) (0.35039596194, 9.024642655) (0.35949204529, 8.992965418) (0.36719964698, 8.954648527) (0.37379776405, 8.927623413) (0.3799693434, 8.894318216) (0.3864117227, 8.859611963) (0.392124378, 8.838728106) (0.3975612344, 8.816682295) (0.4024455886, 8.793602013) (0.4071459387, 8.772307017) 
};
\addlegendentry{complete, Yen-KSP, fittest}
\end{axis}
\end{tikzpicture}}\\

    \subfloat[Network utilization.]{%
      \label{util}%
      \begin{tikzpicture}
\begin{axis}[xlabel = offered load, ylabel style = {align = center},
ylabel = {network utilization},
legend columns = 2,
legend to name = legend,
legend style = {/tikz/every even column/.append style = {column sep = 1 cm}},
height = 4.5 cm, width = 5.8 cm,
every axis y label/.style = {at = {(ticklabel cs:0.5)}, rotate = 90, anchor = near ticklabel},
y tick label style = {/pgf/number format/fixed, /pgf/number format/1000 sep = \thinspace}]
\addplot[solid]
coordinates {
(0.1, 0.0930980107075) (0.15, 0.139769706066) (0.2, 0.187154808557) (0.25, 0.23544713972) (0.3, 0.28295978117) (0.35, 0.32006561615) (0.4, 0.34513922436) (0.45, 0.36267049764) (0.5, 0.37644639512) (0.55, 0.38813917994) (0.6, 0.39767243456) (0.65, 0.40503065699) (0.7, 0.41170651692) (0.8, 0.42380551286) (0.9, 0.43323462252) (1.0, 0.44131654642) (1.1, 0.4484267895) (1.2, 0.45447479961) (1.3, 0.4600625945) (1.4, 0.4653046265) (1.5, 0.4691643869) (1.6, 0.4736731003) (1.7, 0.4775433122) (1.8, 0.4812047) (1.9, 0.4850870407) (2.0, 0.4880015873) 
};
\addlegendentry{proposed, optimal, fittest}
\addplot[dotted, ultra thick]
coordinates {
(0.1, 0.0888911529767) (0.15, 0.133383032563) (0.2, 0.178273208838) (0.25, 0.22411181904) (0.3, 0.27113673088) (0.35, 0.31041925758) (0.4, 0.33688161698) (0.45, 0.35683552133) (0.5, 0.37131461717) (0.55, 0.38258526804) (0.6, 0.3928364426) (0.65, 0.40124901286) (0.7, 0.4087763021) (0.8, 0.42050535919) (0.9, 0.43080108754) (1.0, 0.43879934924) (1.1, 0.44633686656) (1.2, 0.45294031425) (1.3, 0.45874015308) (1.4, 0.46372810715) (1.5, 0.4683725219) (1.6, 0.4725115637) (1.7, 0.4766756751) (1.8, 0.4803704032) (1.9, 0.4840657207) (2.0, 0.4873111937) 
};
\addlegendentry{complete, optimal, fittest}
\addplot[double]
coordinates {
(0.1, 0.0927485985714) (0.15, 0.13662532999) (0.2, 0.174990362087) (0.25, 0.20506157502) (0.3, 0.2269836236) (0.35, 0.24678004574) (0.4, 0.2622578113) (0.45, 0.27515215106) (0.5, 0.28671804033) (0.55, 0.29752427011) (0.6, 0.30692801804) (0.65, 0.3154141886) (0.7, 0.32309091117) (0.8, 0.33625671555) (0.9, 0.3476611073) (1.0, 0.35799279927) (1.1, 0.36688369299) (1.2, 0.37514074417) (1.3, 0.3822672341) (1.4, 0.3897416021) (1.5, 0.3954464001) (1.6, 0.4016719852) (1.7, 0.4069610457) (1.8, 0.411881415) (1.9, 0.4167787224) (2.0, 0.4211295852) 
};
\addlegendentry{proposed, Yen-KSP, fittest}
\addplot[dashed, ultra thick]
coordinates {
(0.1, 0.0887849339883) (0.15, 0.130526366253) (0.2, 0.165881278938) (0.25, 0.19244910243) (0.3, 0.21431485027) (0.35, 0.23238868323) (0.4, 0.24696684074) (0.45, 0.26060904441) (0.5, 0.2717703514) (0.55, 0.28139810489) (0.6, 0.29073166394) (0.65, 0.29904109993) (0.7, 0.30662256818) (0.8, 0.31983074604) (0.9, 0.33199964623) (1.0, 0.34183812008) (1.1, 0.35039596194) (1.2, 0.35949204529) (1.3, 0.36719964698) (1.4, 0.37379776405) (1.5, 0.3799693434) (1.6, 0.3864117227) (1.7, 0.392124378) (1.8, 0.3975612344) (1.9, 0.4024455886) (2.0, 0.4071459387) 
};
\addlegendentry{complete, Yen-KSP, fittest}
\end{axis}
\end{tikzpicture}}&%
    \subfloat[Number of connections.]{%
      \label{conns}%
      \begin{tikzpicture}
\begin{axis}[xlabel = offered load, ylabel style = {align = center},
ylabel = {number of connections},
legend columns = 2,
legend to name = legend,
legend style = {/tikz/every even column/.append style = {column sep = 1 cm}},
height = 4.5 cm, width = 5.8 cm,
every axis y label/.style = {at = {(ticklabel cs:0.5)}, rotate = 90, anchor = near ticklabel},
y tick label style = {/pgf/number format/fixed, /pgf/number format/1000 sep = \thinspace}]
\addplot[solid]
coordinates {
(0.1, 96.4252949495) (0.15, 144.481) (0.2, 192.6066) (0.25, 240.0369) (0.3, 284.5975) (0.35, 323.9094) (0.4, 356.1494) (0.45, 382.7671) (0.5, 407.526) (0.55, 430.7574) (0.6, 450.7255) (0.65, 469.6686) (0.7, 488.005) (0.8, 522.8694) (0.9, 553.5705) (1.0, 582.2195) (1.1, 609.0963) (1.2, 634.5854) (1.3, 658.5593) (1.4, 680.2888) (1.5, 701.3189) (1.6, 722.6537) (1.7, 742.7391) (1.8, 761.8522) (1.9, 781.231) (2.0, 798.2985) 
};
\addlegendentry{proposed, optimal, fittest}
\addplot[dotted, ultra thick]
coordinates {
(0.1, 96.4252949495) (0.15, 144.491) (0.2, 192.7784) (0.25, 240.1844) (0.3, 286.1471) (0.35, 325.81) (0.4, 359.0755) (0.45, 388.4363) (0.5, 413.3031) (0.55, 437.1414) (0.6, 458.4017) (0.65, 477.2466) (0.7, 497.3032) (0.8, 531.7214) (0.9, 563.217) (1.0, 593.1308) (1.1, 619.6259) (1.2, 645.5708) (1.3, 670.7897) (1.4, 692.719) (1.5, 716.7968) (1.6, 736.7331) (1.7, 757.7683) (1.8, 776.1117) (1.9, 796.662) (2.0, 814.0265) 
};
\addlegendentry{complete, optimal, fittest}
\addplot[double]
coordinates {
(0.1, 96.3179949495) (0.15, 142.8043) (0.2, 186.7362) (0.25, 225.5378) (0.3, 257.9162) (0.35, 290.2516) (0.4, 317.4091) (0.45, 341.7972) (0.5, 366.1204) (0.55, 388.4416) (0.6, 409.0436) (0.65, 428.5191) (0.7, 447.7395) (0.8, 483.5839) (0.9, 515.7165) (1.0, 546.9405) (1.1, 574.5929) (1.2, 601.0992) (1.3, 626.6739) (1.4, 652.3919) (1.5, 674.2396) (1.6, 697.67) (1.7, 718.9016) (1.8, 740.0676) (1.9, 759.1363) (2.0, 779.2959) 
};
\addlegendentry{proposed, Yen-KSP, fittest}
\addplot[dashed, ultra thick]
coordinates {
(0.1, 96.2793949495) (0.15, 142.9516) (0.2, 186.2702) (0.25, 222.7281) (0.3, 257.2537) (0.35, 287.5023) (0.4, 314.5412) (0.45, 341.1749) (0.5, 365.3614) (0.55, 386.3958) (0.6, 408.3098) (0.65, 428.9547) (0.7, 447.7289) (0.8, 482.8619) (0.9, 517.1559) (1.0, 547.204) (1.1, 576.4362) (1.2, 604.289) (1.3, 631.6397) (1.4, 656.1506) (1.5, 679.0717) (1.6, 703.2818) (1.7, 725.9895) (1.8, 746.5334) (1.9, 767.6331) (2.0, 787.166) 
};
\addlegendentry{complete, Yen-KSP, fittest}
\end{axis}
\end{tikzpicture}}&%
    \subfloat[Capacity served.]{%
      \label{capser}%
      \begin{tikzpicture}
\begin{axis}[xlabel = offered load, ylabel style = {align = center},
ylabel = {capacity served},
legend columns = 2,
legend to name = legend,
legend style = {/tikz/every even column/.append style = {column sep = 1 cm}},
height = 4.5 cm, width = 5.8 cm,
every axis y label/.style = {at = {(ticklabel cs:0.5)}, rotate = 90, anchor = near ticklabel},
y tick label style = {/pgf/number format/fixed, /pgf/number format/1000 sep = \thinspace}]
\addplot[solid]
coordinates {
(0.1, 601589.113027) (0.15, 906115.6574) (0.2, 1224920.276) (0.25, 1564637.8461) (0.3, 1911944.0415) (0.35, 2191382.504) (0.4, 2379733.7095) (0.45, 2509830.6053) (0.5, 2614345.7217) (0.55, 2701683.6316) (0.6, 2773424.9952) (0.65, 2826639.4387) (0.7, 2876403.3913) (0.8, 2967065.0856) (0.9, 3036166.4113) (1.0, 3095943.1613) (1.1, 3147727.4802) (1.2, 3192161.1461) (1.3, 3233595.1976) (1.4, 3271464.8113) (1.5, 3300066.8341) (1.6, 3333050.2901) (1.7, 3360104.0326) (1.8, 3386924.3444) (1.9, 3415272.7117) (2.0, 3437438.3839) 
};
\addlegendentry{proposed, optimal, fittest}
\addplot[dotted, ultra thick]
coordinates {
(0.1, 559470.341447) (0.15, 841625.6768) (0.2, 1137823.4874) (0.25, 1455669.9801) (0.3, 1796772.5134) (0.35, 2091807.6521) (0.4, 2293226.3343) (0.45, 2443714.8785) (0.5, 2552528.4344) (0.55, 2637550.2262) (0.6, 2713289.017) (0.65, 2776334.3785) (0.7, 2833398.4279) (0.8, 2920523.1466) (0.9, 2996450.0051) (1.0, 3055797.7223) (1.1, 3110111.3069) (1.2, 3159370.4865) (1.3, 3201079.71) (1.4, 3239449.1701) (1.5, 3272191.8399) (1.6, 3303320.453) (1.7, 3332716.1595) (1.8, 3359753.2705) (1.9, 3386397.6495) (2.0, 3410516.7831) 
};
\addlegendentry{complete, optimal, fittest}
\addplot[double]
coordinates {
(0.1, 598295.013884) (0.15, 882784.4731) (0.2, 1137569.9816) (0.25, 1341250.0458) (0.3, 1493614.0664) (0.35, 1630153.2328) (0.4, 1739158.1841) (0.45, 1828999.0423) (0.5, 1911166.2559) (0.55, 1988052.121) (0.6, 2053968.1234) (0.65, 2114331.5461) (0.7, 2168231.2139) (0.8, 2262678.4654) (0.9, 2345398.4794) (1.0, 2419156.0925) (1.1, 2482311.1986) (1.2, 2542707.1308) (1.3, 2594774.3591) (1.4, 2648954.371) (1.5, 2689454.127) (1.6, 2734683.7846) (1.7, 2772132.6681) (1.8, 2810232.5327) (1.9, 2846011.447) (2.0, 2876384.0471) 
};
\addlegendentry{proposed, Yen-KSP, fittest}
\addplot[dashed, ultra thick]
coordinates {
(0.1, 558905.028747) (0.15, 823431.3262) (0.2, 1052819.0929) (0.25, 1229002.5106) (0.3, 1375696.0609) (0.35, 1498394.9166) (0.4, 1599247.4179) (0.45, 1692216.2531) (0.5, 1769040.2468) (0.55, 1836130.7276) (0.6, 1903205.4394) (0.65, 1960569.0099) (0.7, 2013497.6048) (0.8, 2107290.4439) (0.9, 2192064.5404) (1.0, 2262252.8128) (1.1, 2323378.942) (1.2, 2387441.4077) (1.3, 2442876.5645) (1.4, 2492107.4856) (1.5, 2535859.2224) (1.6, 2581174.6609) (1.7, 2623003.6487) (1.8, 2661810.6703) (1.9, 2697478.2352) (2.0, 2732121.7792) 
};
\addlegendentry{complete, Yen-KSP, fittest}
\end{axis}
\end{tikzpicture}}\\[15pt]

  \end{tabular}
  \ref{legend}\\[5pt]
  \caption{Simulation results related to connection and network
  performance.}
  \label{second}
\end{figure*}

Figures \ref{first} and \ref{second} show the sample means of the
measured values as lines, where each line has 26 data points.  The
figures do not report the standard errors as error bars, because they
were too small to be drawn as the aforementioned relative standard
errors were below 1\%.  Most of the sample means are reported as a
function of network utilization, and not of the offered load, because
we are interested in how the reconfiguration algorithms perform under
a given network state expressed by the network utilization.

% A note on ED-ASP - we do not show its results.

We do not plot the obtained simulation results for the ED-ASP routing
algorithm, because they were worse by a few percent than the results
for the Yen-KSP algorithm, and because they would obfuscate the plots.
Our general finding is that Yen-KSP outperforms ED-ASP in establishing
and reconfiguring a connection, and so we compare the results of the
optimal and Yen-KSP routing algorithms only.  The optimal routing
algorithm performs the best.

% The number of links - the proposed algorithm rules.

Of the central interest is the number of new links required for a
reconfigured connection, shown in Fig.~\ref{nonl}.  The proposed
algorithm requires about half as many new links as the complete
reconfiguration, because it reuses more links of the established
connection than the complete reconfiguration, as shown in
Fig.~\ref{nool}.  \emph{The low number of new reconfiguration links is
the reason, we argue, for using the proposed reconfiguration algorithm
over the complete reconfiguration algorithm.}

% The number of links - some details.

The number of all links of a reconfigured connection is a few percent
higher for the proposed configuration than for the complete
reconfiguration as shown in Fig.~\ref{noal}, which we show has
negligible detrimental effect on the network.  The number of links of
a reconfigured connection increases up to the network utilization of
0.35, because the network is able to support circuitous
reconfiguration paths, while for the higher network utilization, the
longer connections are less likely to be reconfigured, and so the
number of all links gets lower.

% The number of links and the spectrum allocation policies.

In Fig.~\ref{first}, for the proposed reconfiguration algorithm, we
show the results for the fittest policy only, because for the first
and random policies the algorithm performed comparably, and plotting
these results would obfuscate the figures.  For the complete
reconfiguration algorithm, we also show the results for the random
policy, since they are the worst: the algorithm does not try to meet
the spectrum continuity constraint, and with the random policy the
algorithm is unlikely to reuse the links of the already established
connection, especially for the lightly utilized network.

% What the next figure shows.

Fig.~\ref{second} shows the results for the fittest policy only,
because for the first and random policies, the results are worse by a
few percent, and reporting them would obfuscate the plots.

% The connection results.

Figures \ref{probs}, \ref{lengths}, and \ref{slices} show the results
for establishing a connection in gray, and for reconfiguring a
connection in black.  We are mainly interested in the reconfiguration
results, and show the results of establishing connections only to
argue that the proposed reconfiguration does not impede the network
performance.

% The probabilities.

The probabilities of establishing and reconfiguring a connection are
shown in Fig.~\ref{probs}.  With the optimal routing algorithm
employed, the proposed reconfiguration algorithm performs equally well
as the complete reconfiguration.

% The lenghts.

The lengths of established and reconfigured connections are shown in
Fig.~\ref{lengths}.  The length of a reconfigured connection is by a
few percent larger for the proposed reconfiguration than for the
complete reconfiguration, analogously to the number of all links of a
reconfigured connection.

% The slices.

The number of slices of a reconfigured connection is shown in
Fig.~\ref{slices}.  As expected, the number of slices gets lower,
since the more slices a connection has, the more difficult it is to
find a candidate reconfiguration path in an increasingly utilized
network.  The number of slices of a reconfigured connection in
comparison with the established connection is lower because of the
reconfiguration case, which requires no slices: the source node is the
new destination node.

% Network utilization.

Fig.~\ref{util} shows the network utilization as a function of the
offered load $\mu$.  The network responds differently to the offered
load depending on the way a connection is established, the way a
connection is reconfigured, and a number of parameters, e.g., $\beta$.
For the optimal routing algorithm, the proposed reconfiguration
algorithm utilizes the network comparably with the complete
reconfiguration.  In comparison with the optimal routing, the Yen-KSP
utilizes the network less, as it is less likely to establish and
reconfigure the connection, and favors short connections with a small
number of slices.

% Connections.

Fig.~\ref{conns} shows the number of established and reconfigured
connections in the network.  The proposed reconfiguration algorithm
perform slightly worse in comparison with the complete
reconfiguration, which is the only detrimental effect of the proposed
algorithm we found.  Even the Yen-KSP routing algorithm performs well
in terms of the number of connections, but its connections are
increasingly shorter as the load increases, as shown in
Fig.~\ref{lengths}.

% Capacity served.

The \emph{capacity served} better captures the network performance, as
it takes into account the length of the connections.  The capacity
served is the sum of the capacities served by every currently
established and reconfigured connection.  The capacity served by a
connection is the product of the number of slices and the path length.
An advantage, albeit minute, is a slightly higher capacity served for
the proposed reconfiguration in comparison with the complete
reconfiguration.

%%%%%%%%%%%%%%%%%%%%%%%%%%%%%%%%%%%%%%%%%%%%%%%%%%%%%%%%%%%%%%%%%%%%%%%%%%%

\section{Conclusion}
\label{conclusion}

We stated a new and important problem of itinerant routing in elastic
optical networks and proposed a novel connection reconfiguration
algorithm.  The algorithm achieves the key objective of minimizing the
number of new links to configure, which is beneficial for real-time
reconfiguration.

We obtained credible simulation results, which suggest the proposed
algorithm performs better than a shortest-path reconfiguration
algorithm.  In comparison with the shortest-path reconfiguration, the
proposed algorithm requires half as many new links to reconfigure, and
performs comparably in terms of several measured metrics related to
establishing and reconfiguring connections, and to the overall network
performance.

For these reasons, we argue, the proposed reconfiguration algorithm is
indispensable for the itinerant routing in the emerging elastic
optical networks, especially in the context of the integration with
the next-generation wireless access networks based on the cognitive or
cloud radio architectures.

Two general findings for both establishing and reconfiguring a
connection are:
\begin{inparaenum}[(a)]
\item the optimal routing algorithm for elastic optical networks
  markedly outperforms the routing with Yen K-shortest paths, and all
  edge-disjoint paths, and
\item the fittest spectrum allocation policy performs slightly better
  than the first and random policies.
\end{inparaenum}

The proposed reconfiguration algorithm could also be used in link
restoration of connections in elastic optical networks, where a
connection rerouting is localized to the end nodes of the failed
optical link.

Future work could concentrate on the network control for itinerant
routing, for example on the trade-offs between the distributed control
(itinerant routing can be coordinated autonomously by any node) and
the centralized control (itinerant routing can be coordinated by a
designated node or a set of nodes).

Further research could elaborate on the proposed proactive calculation
of candidate reconfiguration paths for the foreseeable events, such as
switching a connection to one of the neighboring base stations, or
some selected data centers.

\bibliographystyle{IEEEtran}
\bibliography{all}

\end{document}